\documentclass[pre,aps,10pt,superscriptaddress,twocolumn,floatfix]{revtex4-1}

\usepackage[nice]{nicefrac}
\usepackage{graphicx,color}
\usepackage{amsmath,amssymb,bm,bbm}
\usepackage{amsmath}
\usepackage{braket}
\usepackage{bbold}
\usepackage{float}

\usepackage{subfigure}

\usepackage[plainpages=false,pdfpagelabels,colorlinks=true,linkcolor=red,urlcolor=blue,citecolor=blue,pdftitle={},pdfauthor={},pdfdisplaydoctitle=true,pdfduplex=DuplexFlipLongEdge]{hyperref}

\definecolor{darkred}{rgb}{0.90,0.2,0.2}
\definecolor{darkgreen}{rgb}{0,0.60,.2}
\definecolor{darkblue}{rgb}{0.1,0.3,1}
\definecolor{grey}{cmyk}{0,0,0,0.25}
\definecolor{orange}{cmyk}{0,0.6,0.8,0}

\begin{document}

\title{A detailed study on various phases in dissipative anisotropic Dicke model}
\author{Pragna Das}
\email{Pragna.Das@ijs.si}
\affiliation{Indian Institute of Science Education and Research Bhopal 462066 India}
\affiliation{Department of Theoretical Physics, J. Stefan Institute, SI-1000 Ljubljana, Slovenia}
\author{Saptarshi Saha}
\email{s.saha@tu-berlin.de}
\affiliation{Department of Physical Sciences, Indian Institute of Science
Education and Research Kolkata, Mohanpur - 741 246, WB, India}
\affiliation{Institut für Physik und Astronomie, Technische Universität Berlin, Hardenbergstr. 36, 10623 Berlin, Germany}

\begin{abstract}
We present a comprehensive study of different phases in the Dicke model incorporating both anisotropy and dissipation. We begin with a concise review of the quantum phase transition in this setting, highlighting how these two parameters shift the critical point. We then perform a detailed investigation of the transition from ergodic to nonergodic phases by analyzing the eigenvalue and eigenvector properties of the Liouvillian with the aid of scaling of the Liouvillian gap and the average participation ratio. Our results show that the eigenvector properties of the Liouvillian are consistent with its spectral characteristics, leading to a phase diagram that has similarities with the closed counterpart. Furthermore, we demonstrate that the Liouvillian gap exhibits distinct scaling behaviors in these two phases. Finally, we extend our study to the driven case by applying a Thue–Morse quasiperiodic drive. In this case, we find that bosonic dissipation plays a crucial role in stabilizing the prethermal plateau, offering an effective mechanism to halt the heating effect arising from the quasi-periodic drive.
\end{abstract}

\maketitle 

\section{Introduction}
Exploring many-body quantum dynamics through atom-photon interaction 
is the heart of the atomic, condensed matter, and ultra-cold quantum physics \cite{cohen-tannoudji_atom-photon_1998, leonard2017monitoring, gutzler_lightmatter_2021, RevModPhys.90.031002}. Experiments using 
cavity QED setup provide an ideal platform to investigate the underlying 
microscopic properties, e.g., the collective effects, various phase 
transitions, Bose-Einstein condensation, etc \cite{walther_cavity_2006, RevModPhys.87.1379, PhysRevX.8.011002, dimer2007proposed}. From a theoretical perspective, the Dicke model (DM) 
is an iconic model for understanding the above aspects of such systems \cite{garraway_dicke_2011, kirton2019introduction,villaseñor2024}. 

The study of collective light–matter interactions traces back to the pioneering work of Dicke, which describes the interaction between 
$N$ ensemble of atoms 
with a single-mode bosonic field via a dipolar coupling strength ~\cite{dicke1954coherence, lambert2004entanglement,
emary2003chaos, emary2003quantum}. In 
the thermodynamic limit ($N\to\infty$), the model shows a quantum phase 
transition (QPT) from normal to superradiant phase~\cite{dicke1954coherence, chavez2016classical, kirton2018superradiant, kirton2019introduction, das2022revisiting}.  Along with it, such a model also shows the excited state phase transition (ESQPT) and thermal phase transition (TPT) under different circumstances. First experimental proposal of observing  the Dicke quantum phase transition using multilevel atoms and cavity-mediated
Raman transitions were proposed by  Dimer \emph{et\ al.} \cite{Dimer2007}. Around the same time, Baumann et al. observed the Dicke phase transition in an open system consisting of a Bose-Einstein condensate coupled to an optical cavity, highlighting the emergence of a self-organized supersolid phase \cite{baumann_dicke_2010}. Such studies later extended to observe Higgs and Goldstone modes in a supersolid quantum gas \cite{leonard_monitoring_2017}.

 A dissipative realization of the generalized Dicke model was also introduced to observe the effects from the counter-rotating terms, and the effect of external drive was further explored \cite{Guti2018}.  The generalized version of 
the Dicke model, as mentioned above, namely the anisotropic Dicke model (ADM)~\cite{buijsman2017nonergodicity, kloc2017quantum, gutierrez2018dissipative, shapiro2020universal, das2023phase}, where the 
coupling strengths correspond to the rotating and counter-rotating 
terms are considered to be different. Due to the asymmetry in the coupling 
strengths, ADM shows novel features in addition to the well-known existing properties, i.e., ergodic to non-ergodic transition (ENET)~\cite{buijsman2017nonergodicity, hu2021out}.  There exists
a one-to-one correspondence between the non-ergodic phase to the integrability of the system. Hence, a detailed study on the non-ergodic phase will be useful in understanding the following emerging phenomena, e.g., many-body localization, quantum scars, prethermalization, etc \cite{RevModPhys.83.863, Mori_2018, serbyn_quantum_2021}. On the other hand, the ergodic phase connects to the non-integrable phase, a path to study the quantum chaos \cite{dalessio_quantum_2016, BORGONOVI20161}. A natural question, therefore, concerns how the aforementioned phases behave in the presence of dissipation, and in particular whether dissipation destroys their non-ergodic character.

In most cases, a leaky cavity is considered to introduce the dissipative effects on the atom-photon interaction \cite{PhysRevA.85.013817, kirton2019introduction,PhysRevLett.118.123602, kirton2018superradiant}. In the case of QPT, the effect of dissipation is previously explored \cite{kirton2018superradiant}. It is noteworthy to study the impact of dissipation in the ENET, which we explore in this manuscript. The Hamiltonian approach is insufficient to capture such dissipative dynamics. Hence, we adapt the Liouvillian approach motivated by the Lindblad quantum master equation to incorporate the non-unitary dynamics \cite{breuer2002theory}. 
In this prescription, the Liouvillian gap ($\Delta$) plays an important role in characterizing the asymptotic convergence to the steady 
state~\cite{haga2021liouvillian, zhou2022exponential,prosen2008quantum, 
kessler2012dissipative, cai2013algebraic, kastoryano2013rapid, 
vznidarivc2015relaxation, casteels2017critical, minganti2018spectral, 
shibata2019dissipative, shibata2019dissipative2, shibata2020quantum, 
mori2020resolving, mori2023symmetrized, 
shirai2023accelerated, mori2023liouvillian, landi2022nonequilibrium,  yuan2021solving}. The dependence of such a Liouvillian gap on the system size for various systems has already been studied, where the nature of the dissipation and the corresponding phases determine such a scaling \cite{zhou2022exponential,prosen2008quantum, vznidarivc2015relaxation, 
shibata2020quantum, mori2020resolving}.  Based on these results, one can conjecture that the length dependence of the Liouvillian gap is also sensitive to the specific phases of many-body dissipative systems. Motivated by this perspective, we aim to investigate the system-size dependence of the Liouvillian gap for ENET in the ADM.
Similarly, the level statistics of the eigenvalues of the Liouvillian provide useful information regarding the phases. For example, in the case of ADM, the system follows the $2D$ Poisson statistics in the non-ergodic phase, whereas in the ergodic phase the system behaves as the Ginibre ensemble~\cite{ginibre1965statistical, grobe1988quantum, markum1999non, haake1991quantum, hamazaki2020universality}. For completeness of the above analysis, the distribution of the eigenvector is further required. We note that for non-Hermitian systems, the correspondence between the eigenvalue-eigenvector property is quite different from its Hermitian counterpart. Motivated by this fact, we performed the study of the average participation ratio of dissipative ADM to provide further explanation for the emergence of ENET.

Another important aspect of a many-body quantum system is exploring the dynamics under an external drive \cite{RevModPhys.89.011004}. In such cases, it is expected that the system will reach the infinite temperature state (e.g., thermal state) \cite{RevModPhys.91.021001, nandkishore_many-body_2015}. An essential distinction can be observed for the periodic drive, where instead of thermalization,  a prethermal quasi-steady state emerges in the intermediate time scale \cite{PhysRevLett.115.256803, PhysRevLett.127.170603, PhysRevA.107.022206, das2023periodically}. Using the Floquet theory, an effective Hamiltonian can be constructed for the prethermal state in the high-frequency limit, and the relevant conserved quantities can be derived subsequently \cite{PhysRevB.103.L140302}. More recently, research interest has shifted towards exploring quantum dynamics under quasi-periodic drive \cite{nandy2017aperiodically, 
mukherjee2020restoring, zhao2021random, zhao2022localization, das2023periodically, tiwari2024dynamical, tiwari2024periodically}, where it is expected that the lifetime of the prethermal phase is shortened due to the absence of temporal order. As such, the quasi-periodic drive further pushes it to the thermal state.  A similar study was also reported in our previous work of ADM \cite{das2023periodically}.  A further question concerns what happens in the presence of dissipation. We want to specify the reason why such a study is important. In the case of dissipative ADM, we consider the spontaneous emission of the cavity modes. Therefore, the cavity and the atoms try to reach their respective ground state in the presence of strong dissipation. The ground state has the lowest entanglement entropy (EE). At the same point, the quasi-periodic drive increases the EE due to heating. We expect a competition between the strength of the quasi-periodic drive and the dissipation rate in this system. We mostly focus on whether the strong dissipation stabilizes the prethermal phase or the quasi-periodic drive dominates the heating process. We also show that such a mechanism is different for the ergodic and non-ergodic phases.
 
As stated above, the primary motivation of this manuscript is to investigate the impact of dissipation on the emergent phases of the ADM in both undriven and driven scenarios. To this end, we outline the structure and main results of the paper as follows. In Sec .~\ref {sec-II}, we introduce the dissipative ADM, where a spin–boson system is coupled to an additional bosonic environment.
We begin with the undriven case in Sec.~\ref {sec-iii}. The existence of QPT is analyzed in Sec.~\ref {sec-iii-A}, followed by a detailed study of the ENET in Sec.~\ref {sec-iii-B}. In particular, we examine the Liouvillian gap in Sec.~\ref {sec-iii-B-a} and investigate the properties of the corresponding eigenvectors in Sec.~\ref {sec-iii-B-b}.
We then turn to the driven case in Sec.~\ref {sec-iv}, where the quasi-periodic driving protocol is defined in Sec.~\ref {sec-iv-a}, and the role of dissipation in the presence of drive is discussed in Sec.~\ref {sec-iv-b}. Finally, we summarize our findings and present a broader perspective on their implications in Sec.~\ref {sec-v}.

\section{Description of the system}
\label{sec-II}
Here we consider a system consisting of $N$ non-interacting two-level atoms coupled with a single-mode bosonic leaky cavity.
    The corresponding Hamiltonian of the system is:
    \begin{eqnarray}
    \mathcal{H} &=& \omega a^{\dagger}a + \omega_0 J_z + \frac{g_1}{\sqrt{N}}( a^{\dagger} J_- + a J_{+}) +\nonumber\\ && \frac{g_2}{\sqrt{N}}( a^{\dagger} J_+ + a J_{-}).
    \label{eqn:hamiltonian_ADM}
    \end{eqnarray}
    where $g_1$ and $g_2$ are the corresponding rotating and counter-rotating spin-boson 
    coupling strength. In general, $g_1 \neq g_2$ indicates the anisotropy in the Dicke model~\cite{das2023phase}, and $g_1=g_2=g$ recovers the original isotropic Dicke model.
    Here $a$ ($a^{\dagger}$) is the bosonic annihilation (creation) 
    operator, and $J_{\alpha}=\sum_{i=1}^{N}\frac{1}{2}\sigma_{\alpha}^{i}$ 
    ($\alpha=$ $x$, $y$ $z$), total angular momentum of the ensemble of 
    $N$ atoms. $\omega$ is the bosonic mode frequency and $\omega_0$ 
 is the level splitting of the atoms.
 The first two terms represent the free Hamiltonian of the bosonic cavity 
 and atoms, respectively.
 The basis of the Hamiltonian is: $\vert n, m\rangle$, where 
    $n=0,1,2...,n_{\text{max}}$: boson number and for the symmetric spin 
    subspace considering even $J$, $m=-J, -J+1,...,0,..., J-1, J$. Thus, the Hilbert space dimension of the closed system is: $N_{\text{D}} 
    = (N + 1)(n_{\text{max}} + 1)$.

 We assume that the spontaneous emission rate of the leaky cavity is $\kappa$ and 
 all other sources of dissipation and dephasing are excluded in 
 our model.  The 
 dynamics 
 can be well explained by the Lindblad master equation in the Markovian regime \cite{breuer2002theory}. 
 The dynamical equation of atom+cavity is given by:  
 \begin{equation}
 \frac{d\rho}{dt} = \mathcal{L}\rho,
    \label{eqn:lindblad}
    \end{equation}
 where $\rho$ is density matrix of the system and $\mathcal{L}$ 
 is the Liouvillian super-operator. 
 The explicit form of $\mathcal{L}\rho$ is:
 \begin{equation}
    \mathcal{L}\rho = -i[H, \rho] + \kappa[2a\rho a^{\dagger} - \{ a^{\dagger}a, \rho \}].
    \label{eqn:lindblad}
    \end{equation}  
 The corresponding analytical form of $\mathcal{L}$ is given by:
  \begin{eqnarray}
    \mathcal{L} &=& -i\left[ (\mathcal{H}\otimes\mathcal{I}) - (\mathcal{I}\otimes\mathcal{H}^{T}) \right] + \kappa\Big( 2a\otimes a^{*} - a^{\dagger}a\otimes\mathcal{I}\nonumber\\
    && - \mathcal{I}\otimes a^{\dagger}a \Big).
    \label{eqn:liouvillian}
    \end{eqnarray}
    The basis 
    of the Liouvillian is $\vert n_l, m_l\rangle\otimes\vert n_r, m_r
    \rangle$ and the corresponding  Liouvillian space dimension is $N_{\text{L}} =
    N_{\text{D}}\times N_{\text{D}}$. Using parity symmetry, the dimension of 
    the Liouvillian space is reduced to $\frac{N_{ \text{ L}}}{2}$~\cite{prasad2022dissipative}. 
In such systems, we are primarily interested in the following phases, as summarized below.
\begin{enumerate}
    \item Absence of external drive,
    \begin{enumerate}
        \item Normal and Superradiant phases
        \item Ergodic and non-Ergodic phases
    \end{enumerate}
    \item Presence of external drive, 
    \begin{enumerate}
        \item Prethermal phase and further heating 
    \end{enumerate}
\end{enumerate}
In the remainder of the manuscript, we will briefly discuss the existence of such phases in ADM.

\section{Analysis for the undriven case}
\label{sec-iii}
We begin with the undriven case, where we aim to study the effects of dissipation on QPT and ENET.

\subsection{Emergence of quantum phase transition (QPT)}
\label{sec-iii-A}
As QPT is well defined and has been discussed thoroughly in previous works, we provide a concise review of these phenomena. Let us consider an isotropic Dicke model, $g_1 =  g_2$. The Hamiltonian of the system is invariant under the following transformation, $a \to -a$, $\sigma_x \to - \sigma_x$ (i.e., $\mathbb{Z}_2$ symmetry). QPT is defined in this case as the breaking of $\mathbb{Z}_2$ symmetry. As such, the normal phase is the $\mathbb{Z}_2$ protected phase, with $\langle a \rangle = 0$, and in the case of a symmetry-broken superradiant phase, $\langle a \rangle \neq 0$ \cite{kirton2019introduction}. Following the retarded Green’s function and the Keldysh path integral formalism, the critical coupling strength for the second-order QPT is defined as $g_c=\frac{1}{2}\sqrt{\omega\omega_0}\sqrt{1+\kappa^2/\omega^2}$. Following the same procedure,  in case of ADM, instead of a line, we  get a region of $g_1$ and $g_2$, which satisfy the following relation,
\begin{equation}
   \left( g_1^2 - g_2^2 \right)^2 - 2\left( g_1^2 + g_2^2 \right)\omega\omega_0 + \left( \omega^2 + \kappa^2 \right)\omega_0^2 = 0,
   \end{equation}
Several experimental platforms have been developed to observe QPT in the Dicke model \cite{Dimer2007,baumann_dicke_2010,leonard_monitoring_2017}.

\subsection{Emergence of ergodic to non-ergodic phase transition (ENET)}
\label{sec-iii-B}
In this section, we provide a concise review of the ENET in closed ADM.  In case of ADM, the system is integrable for $g_1 =0$ or $g_2 = 0$. As a consequence, the system shows ENET in the vicinity of  the integrable regime. Previous work shows the emergence of ENET in closed ADM with the aid of level spacing ratio and also shows that such a phase transition is not the precursor of QPT \cite{buijsman2017nonergodicity}. We will further extend this approach to the dissipative case and study the properties of the Liouvillian of such systems for different values of $g_1$ and $ g_2$.

\subsubsection{Finite size scaling of the Liouvillian gap}
\label{sec-iii-B-a}
\begin{figure*}[t] 
  \subfigure{\includegraphics[width=0.45\textwidth]{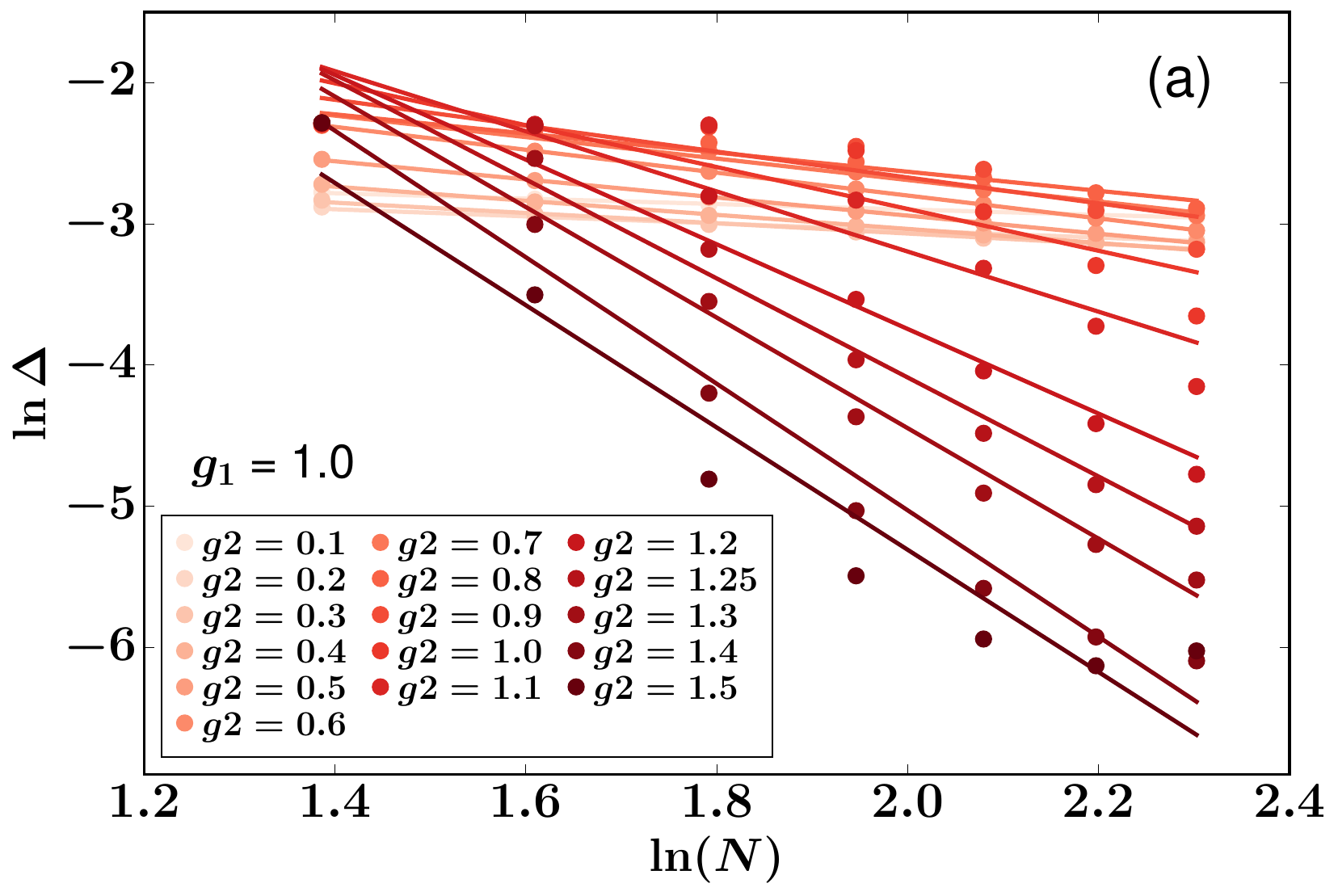}}
  \subfigure{\includegraphics[width=0.45\textwidth]{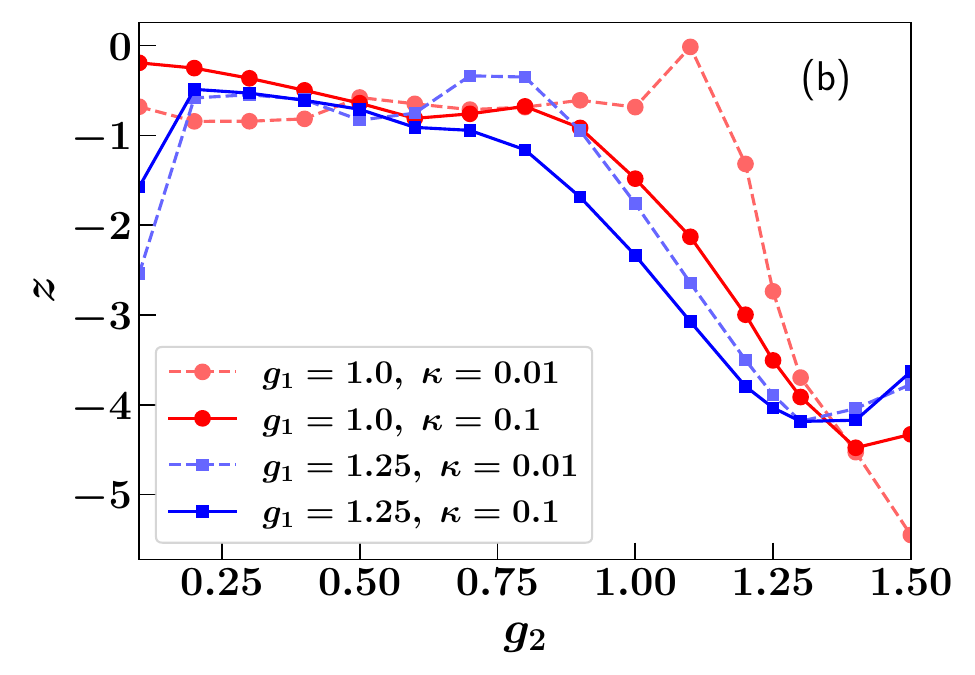}}
  \caption{(a) $\ln\Delta$ vs $\ln N$ plot for $g_1=1.0$, and $\kappa=0.1$. (b) The slope $z$ (extracted from the scaling of Liouvillian gap $\Delta$ as a function of atom number) as a function of $g_2$ for fixed $g_1$ value. We consider the dissipation strength, $\kappa=0.1, 0.01$. Here, $n_{\text{max}}=26$, and the atom number $N$ is changing from $4-10$. We find that the scaling is different for the ergodic and non-ergodic phases. We also show that for changing $\kappa$, the qualitative features of ENET remain the same. }
  \label{fig:L_gap1}
\end{figure*}

An analysis using the eigenvalues of the Liouvillian ($\mathcal{L}$) is required to understand ENET in ADM. The real part of the eigenvalues of  $\mathcal{L}$ is always less than or equal to zero. The real zero eigenvalue shows the notion of the steady state. On the other hand, the negative real parts are the decaying modes. If the complex part is non-zero, then such eigenmodes show  oscillatory behavior \cite{saha2024prethermalization, kessler2012dissipative}. Whether such oscillation is long-lived or not is decided by the corresponding real part of the eigenvalue.
The Liouvillian 
    gap, $\Delta$, is defined by the real distance of the nearest 
    neighbor eigenvalue of $\mathcal{L}$ from its zero eigenvalue. 
\begin{equation}
    \Delta = \text{min}[\text{Re}(E_0-E_n)],
    \label{eqn:gap}
    \end{equation} 
    where $E_n$ are the eigenvalues of the liouvillian and $E_0 = 0 + 0i$, 
    is the corresponding minimum eigenvalue.  The inverse of the Liouvillian gap, $1/\Delta$, shows the notion of the relaxation time of the system.
   In Fig.~\ref{fig:L_gap1}(a), we plot $\ln \Delta$ versus $\ln N$ for fixed $g_1 = 1.0$ and $\kappa = 0.1$, and for different values of $g_2$ ($0.1 \leqslant g_2 \leqslant 1.5$). From the linear fits in Fig.~\ref{fig:L_gap1}(a), we extract the scaling exponent $z$ defined by $\Delta \propto N^{z}$. The resulting dependence of $z$ on $g_2$ is shown in Fig.~\ref{fig:L_gap1}(b). We find that the slope of $z$ is different for the non-ergodic phase and the ergodic phase, respectively. The system exhibits a non-ergodic response when $g_2$ is small, while increasing $g_2$, we find a crossover into the ergodic phase. A more thorough analysis is required to reliably identify the phase boundary of ENET, in a manner similar to that used for characterizing a QPT. 

A true phase transition would require taking the thermodynamic limit $N \to \infty$; however, the high dimensionality of the system makes this limit challenging to access. Hence, our result shows a phase crossover between the ergodic and non-ergodic phases rather than a phase transition. 

We check the same for $g_1 =1.0, \, 1.25$ and $\kappa=0.1,\ 0.01$ in Fig.~\ref{fig:L_gap1}(b), and our result shows that for changing $\kappa$, the system shows qualitatively similar response.  
We also provide a contour plot of the exponent $z$ by varying $g_1$ and $g_2$ in Appendix  \ref{app3} to provide other details on ENET.

\subsubsection{Average participation ratio using eigenvector properties}
\label{sec-iii-B-b}
The eigenvalue properties of the Liouvillian for the dissipative Dicke model are well studied  previously ~\cite{prasad2022dissipative}. A recent study including  the anisotropic case of this particular model also exists ~\cite{villasenor2024breakdown}. However, for completeness, a study requires both the eigenvalue and the eigenvector properties to understand the phase diagram of this model. We also note that for non-Hermitian systems, the correspondence  between the eigenvalue and eigenvector is not as straightforward as in the Hermitian case. For example, if the Liouvillian of the dissipative system has a non-zero real eigenvalue, then the corresponding eigenvector is traceless, which is not true for the Hamiltonian of the Hermitian case. This observation motivates us to  study the eigenvector properties of the dissipative ADM in order to gain deeper insight into ENET.

In general, Liouvillian is a non-Hermitian matrix with complex eigenvalues, $\mathcal{L} \rho_i = \lambda_i \rho_i$. If $\mathcal{L}$ is diagonalizable, then using the spectral decomposition theory, any operator $\hat{A}$ in the Liouville space can be written as $\hat{A} = \sum c_i \rho_i$. The value of $c_i$ can be calculated from the standard orthogonality relation for a non-Hermitian matrix, $Tr(\Theta_j^\dagger \rho_i) = \delta_{ij}$. Here, $\Theta_j$  is the left eigenvector of $\mathcal{L}$, i.e., $\mathcal{L}^\dagger \Theta_i = \lambda_i \Theta_i$. The above definition implies that to calculate the participation ratio $PR \propto \frac{1}{\sum_{i=1}^N \vert c_i \vert^4} $~\cite{das2023phase}, we must use a bi-orthogonal basis consisting of both left and right eigenvectors.
We note that the results corresponding only to the left and only to the right are not quite correct and may provide wrong results.

 To calculate the participation ratio, in the Liouville space, we perform the LU decomposition for bi-orthogonalization. Let $L$ and $R$ be the corresponding left and right eigenvectors of $\mathcal{L}$, and we define $L^{\dagger}R=M$.  The matrix $M$ can be represented in its LU-decomposed form: $M = 
M_{\text{L}}M_{\text{U}}$, where $M_{\text{L}}(M_{\text{U}})$ is the 
lower(upper) triangular matrix. Hence,
\begin{equation}
L^{\dagger}R = M_{\text{L}}M_{\text{U}} 
\end{equation}
Following the above approach, one can redefine the left and right eigenvectors as $ L
\rightarrow L \left(M_{\text{L}}^{-1}\right)^{\dagger}$ and $R
\rightarrow R M_{\text{U}}^{-1}$ such that $L^{\dagger} R = 1$ to make the Liouvillian space bi-orthogonal.
The definition of PR is given by:
\begin{equation}
P^{\rm LR}_{n} = \frac{\left( \sum_{i}|\tilde{\psi}_i^n| \right)^2}{\sum_{i}|\tilde{\psi}_i^n|^2}.
\end{equation}
The superscript $\rm LR$ represents biorthogonal PR. 
$\tilde{\psi}_i^n = (\langle L_{n}|b_i\rangle^{*})\langle R_{n}|b_i\rangle$ 
with $R_n$ and $L_n$ are the $n^{\text{th}}$ right and 
left eigenstates respectively, and $\vert b_i \rangle$'s are the Fock space chosen as 
computational basis. The average PR is defined as:  
$P^{\rm LR}_{\rm avg}=\sum_n P^{\rm LR}_{n}/N_{\rm L}$. 
We check that such a result is also consistent without the $LU$ decomposition.
\begin{figure}
\subfigure{\includegraphics[width=0.5\textwidth]{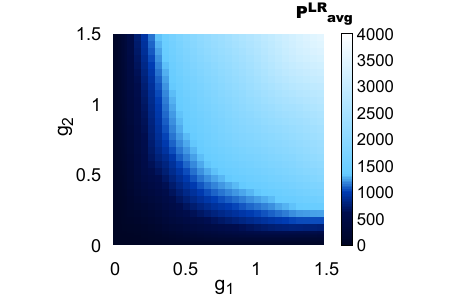}\label{fig:ravg_g_1pt0}}
\caption{Average of participation ratio ($P_{\text{avg}}$), using biorthonormal basis considering both left
and right eigenvectors of the Liouvillian for dissipative ADM  as a
function of the rotating and counter-rotating coupling parameters $g_1$
and $g_2$. The atom number: $N=10$, the bosonic cut-off: $n_{\text{max}}
=26$. For thisfigure $\omega_0 = \omega=1,\ \kappa=1$.  }
\label{fig:PR_avg}
\end{figure}
In Fig.~\ref{fig:PR_avg}, we show the value of
$P_{\rm avg}^{\rm LR}$ as a function of the coupling strengths. In the dark portion, the average participation ratio is
relatively low, which indicates the non-ergodic phase, compared to the
light color regime, where the average participation is quite high and
represents the ergodic phase. The phase diagram is similar to
Fig.~\ref{fig:rav_OADM} in the Appendix- \ref{app}, where we show the transition from the
non-ergodic to the ergodic phase with the aid of the level spacing ratio of eigenvalues of $\mathcal{L}$. Hence, we conclude that the
eigenvector properties of our system are consistent
with the eigenvalue properties. Our result also indicates that dissipation does not destroy the non-erogodicity in the system in the vicinity of $g_1 = 0$ or $g_2 = 0$. We also provide supporting plots in Appendix \ref{app2} to demonstrate that the qualitative features remain unchanged as the dissipation rate $\kappa$ is varied.
    
\begin{figure*}
    \subfigure{\includegraphics[width=0.48\textwidth]{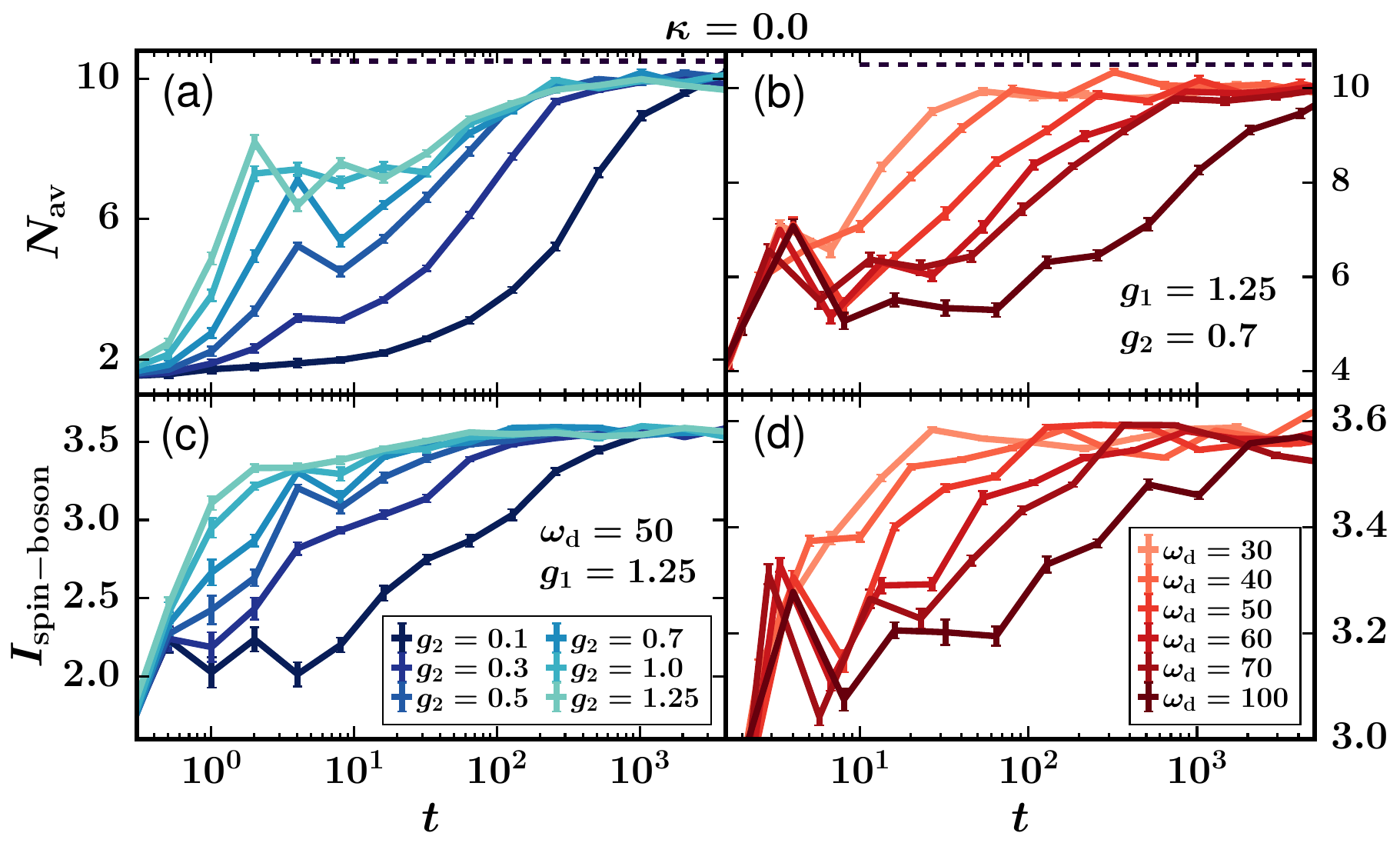}\label{fig:ravg_g_0pt2}}
    \subfigure{\includegraphics[width=0.48\textwidth]{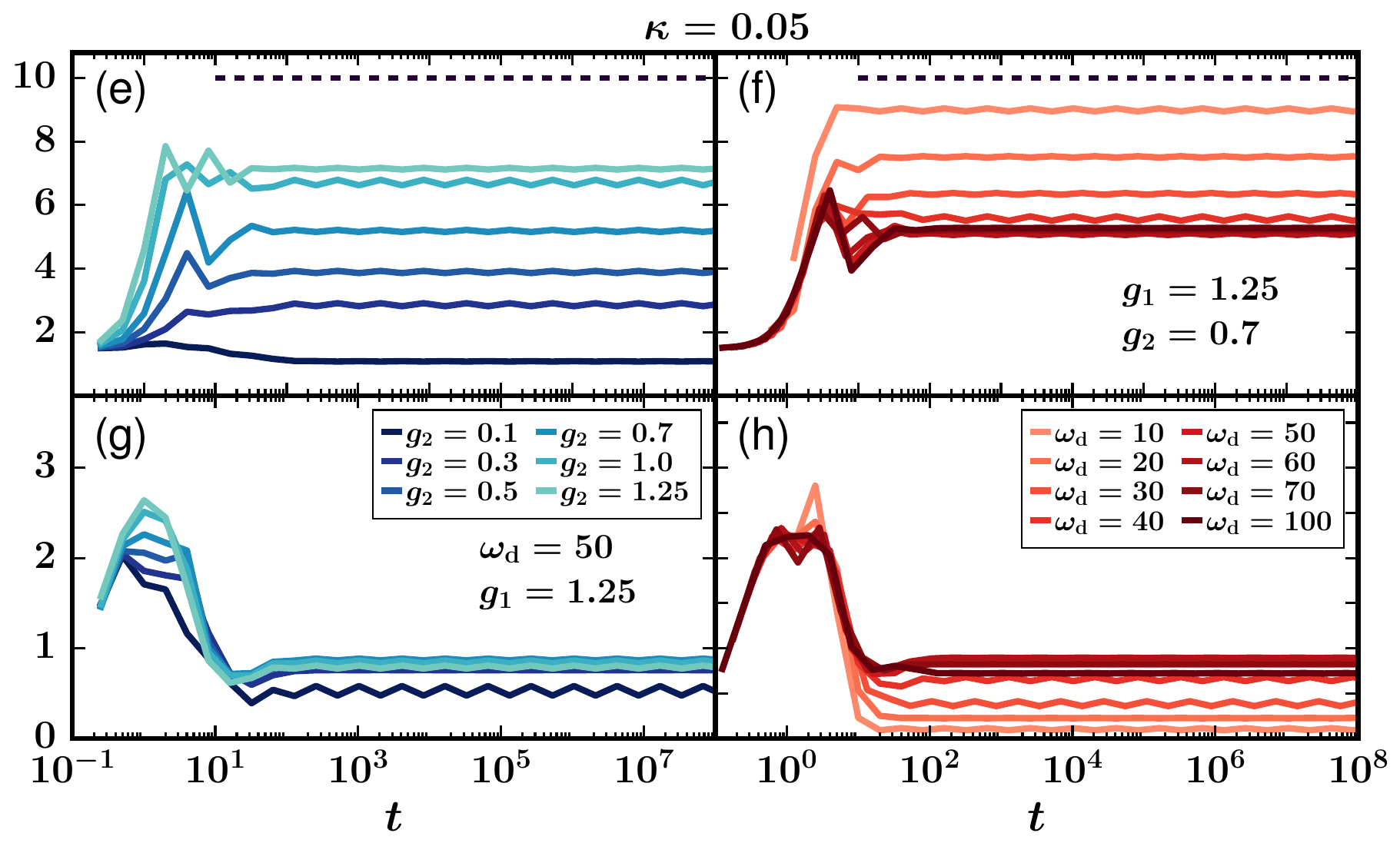}\label{fig:ravg_g_0pt2}}
    \subfigure{\includegraphics[width=0.48\textwidth]{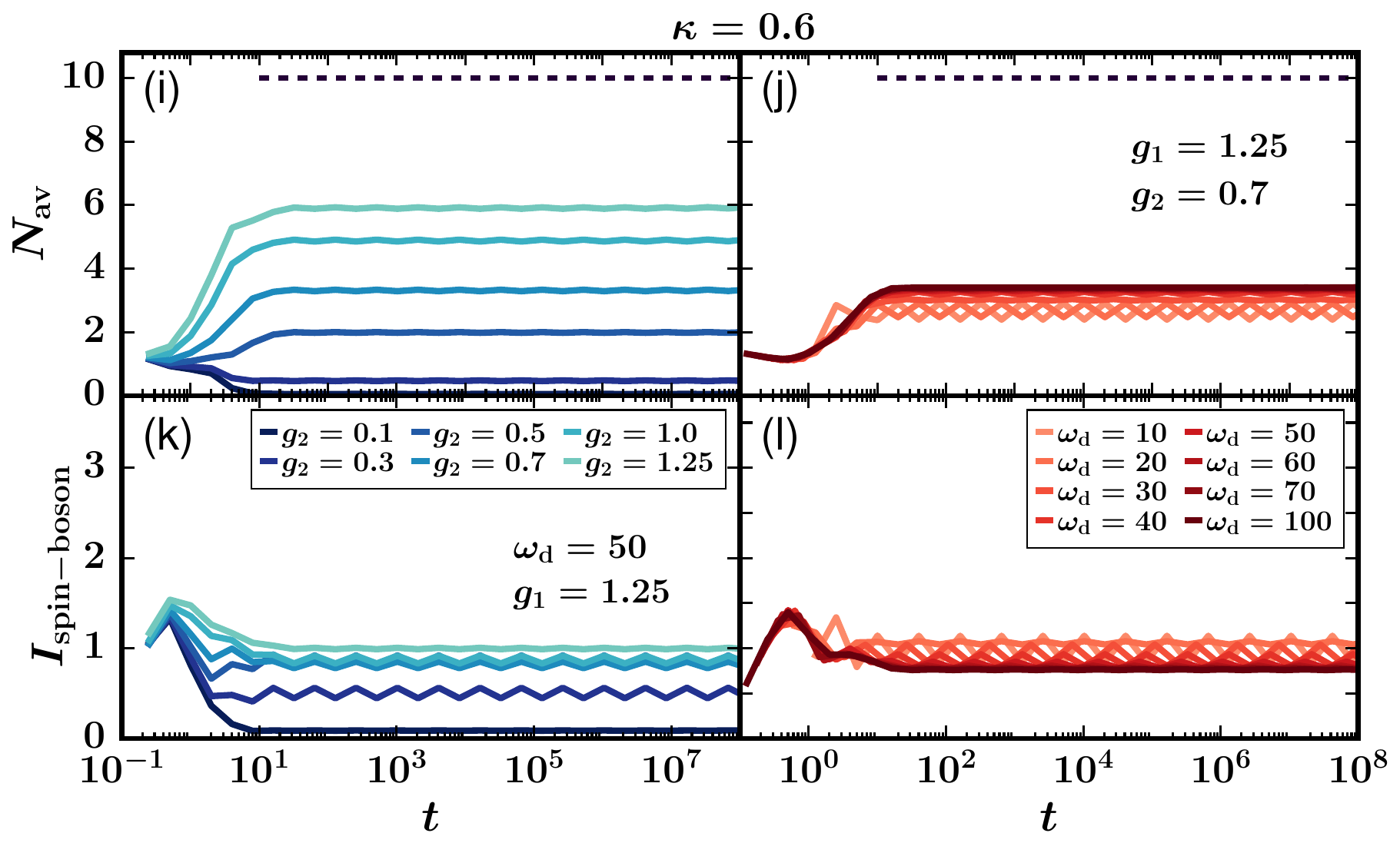}\label{fig:ravg_g_0pt2}}
    \subfigure{\includegraphics[width=0.48\textwidth]{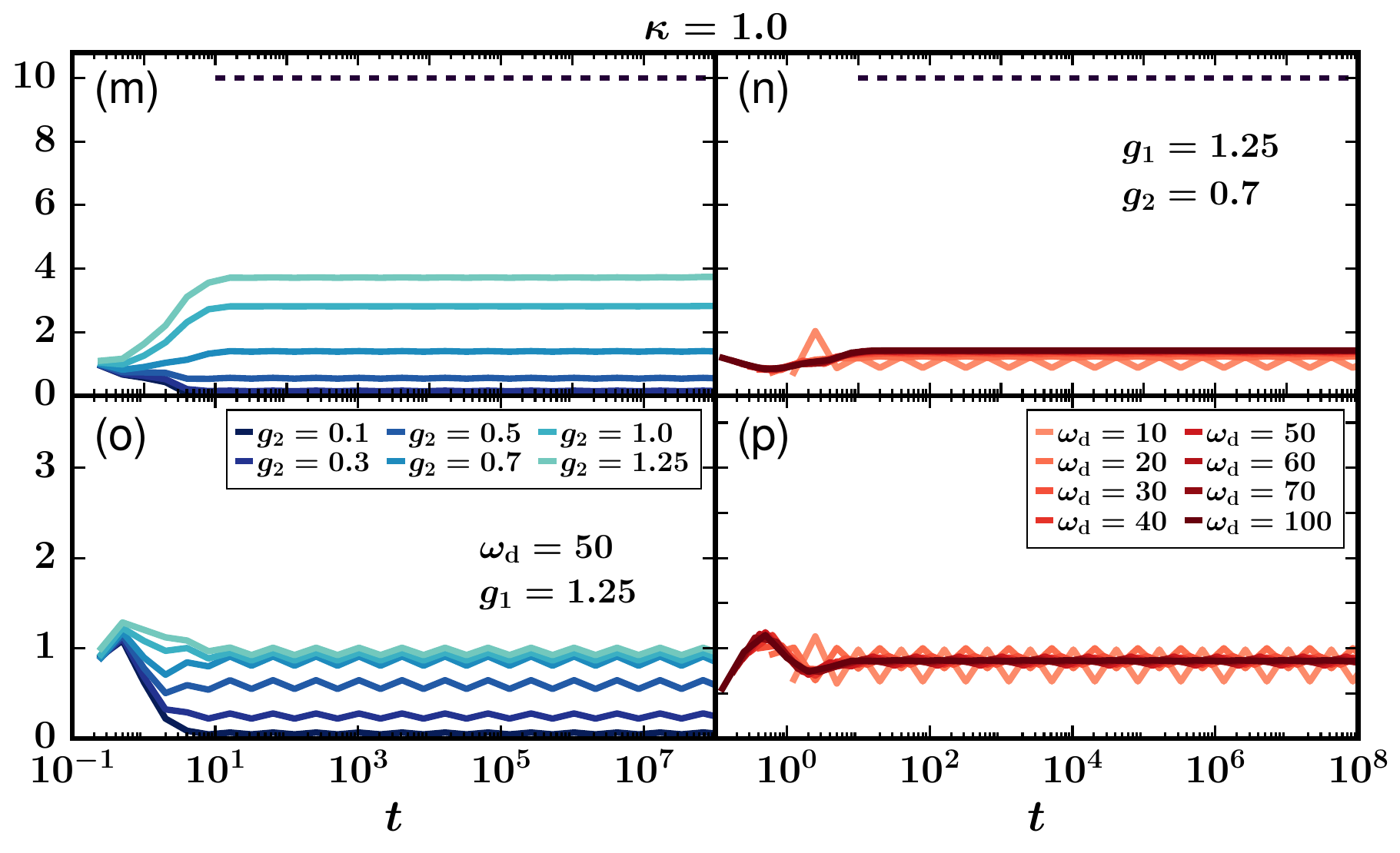}\label{fig:ravg_g_0pt2}}
    \caption{(a, b, e, f, i, j, m, n) Average boson number, $N_\text{av}(t)$, 
    and (c, d, g, h, k, l, o, p) mutual information between spins and 
    bosons, $I_{\text{spin-boson}}$, as a function of time $t_n=2^n T$ 
    for the dissipative ADM under the Thue-Morse 
    quasiperiodic drive. The initial states have low energies, so that 
    $\langle E_{\text{in}}\rangle = 0.25$. Panels (a, c, e, g, i, k, 
    m, o) represent the data for a fixed driving frequency $\omega_{
    \text{d}} = 50$, $g_1 = 1.25$, and various values of $g_2$. Panels 
    (b, d, f, h, j, l, n, p): $g_1 = 1.25$, $g_2 = 0.7$ 
    and various 
    values $\omega_{\text{d}}$. In all panels, the driving amplitude 
    is $\Omega=1.0$, $N=6,\ n_{\text{max}}=20,\ \omega_0 = \omega=1$. 
    For (a-d) $\kappa=0.0$, (e-h) $\kappa=0.05$, (i-l) $\kappa=0.6$, 
    (m-p) $\kappa=1.0$. The time t is in units of $\omega_0^{-1}$. In this figure, the dashed line represents the
page value}
  \label{fig:qd_TM}
\end{figure*}   
\section{Analysis for the driven case}
\label{sec-iv}
In our earlier study of the closed version of driven ADM~\cite{das2023periodically},
we show that under the external periodic drive, the system reaches a long-lived prethermal plateau that escapes Floquet heating. In contrast, for the quasi-periodic drive, the system features a short-lived prethermal plateau, and further heating is observed. At the same time, the dynamics shows the following properties,
\begin{enumerate}
\item With increasing the driving frequency, the heating process becomes slow.
\item In the non-ergodic regime, for the same driving frequency, heating occurs more slowly than in the ergodic regime.
\end{enumerate}
Such analysis can also be extended to the dissipative case.
In case of strong bosonic dissipation, the system will reach ground state  where the mutual information between the spins and boson is zero.
Whereas the quasiperiodic drive pushes the system to a state with higher mutual information. Hence, a competition between the quasiperiodic drive and the bosonic dissipation can be expected in this case.

 \subsection{Driving protocol}
 \label{sec-iv-a}
 We briefly discuss the quasi-periodic 
driving protocol. In particular, we consider the quasi-periodic 
protocol consisting of the Thue-Morse sequence. 
Here, we define two Hamiltonian $H_{\pm}= \omega a^{\dagger}a 
+ \omega_0 J_z + \frac{(g_1 \pm \Omega)}{\sqrt{N}}(a J_+ + a^{\dagger} J_-) 
+ \frac{(g_2 \pm \Omega)}{\sqrt{N}}(a J_- + a^{\dagger} J_+)$, where 
$\Omega$ is the driving amplitude and the corresponding Liouvillian 
using Eq.~\eqref{eqn:liouvillian} are $\mathcal{L}_{\pm}$. The 
Thue-Morse sequence~\cite{thue1906uber, nandy2017aperiodically, 
mukherjee2020restoring, zhao2021random, zhao2022localization} can be 
constructed by starting time evolution operators $\mathcal{U}_{\pm}=
\exp(\mathcal{L}_{\pm}T)$. We can build up the next sequence using 
$\mathcal{U}_1 = \mathcal{U}_-\mathcal{U}_+$ and $\tilde{\mathcal{U}}_1 
= \mathcal{U}_+\mathcal{U}_-$, $\mathcal{U}_2 = \tilde{\mathcal{U}}_1 
\mathcal{U}_1$ and $\tilde{\mathcal{U}}_2 = \mathcal{U}_1 \tilde{\mathcal{U}}_1$  
and so on. The driving unit cells of time length $2^n T$ can be recursively constructed as: $\mathcal{U}_{n+1} = \tilde{\mathcal{U}}_n 
\mathcal{U}_n,\   \tilde{\mathcal{U}}_{n+1} = \mathcal{U}_n \tilde{\mathcal{U}}_n$.  
The time evolution of the initial density matrix is given by 
$\rho_{n}|_{N_{\text{L}}\times 1} = \rho(t = 2^n T)|_{N_{\text{L}}\times 1} = 
\mathcal{U}_n|_{N_{\text{L}}\times N_{\text{L}}}\times \rho_{\text{in}}|_{N_{\text{L}}\times 1}$, 
where $\rho_{\text{in}}|_{N_{\text{L}}\times 1} = 
|\psi_{\text{in}}\rangle|_{N_{\text{D}}\times 1}\otimes|\psi_{\text{in}}\rangle|_{N_{\text{D}}\times 1}$ 
is the initial density matrix and $|\psi_{\text{in}}\rangle$ is the 
initial state.  We also define $\omega_d = 2\pi/T$. To calculate the expectation value of an observable, we 
reshape the density matrix as: 
$\rho(t)|_{N_{\text{L}}\times 1}\rightarrow \rho(t)|_{N_{\text{D}}\times N_{\text{D}}}$.
Here we take the average over $20$ low-energy eigenstates 
(starting from the ground state) of the decoupled Hamiltonian ($g_1=g_2=0.0$).

\subsection{Effect of the dissipation in Floquet heating}
\label{sec-iv-b}
To study quantum dynamics, we define the following relevant observables: the average boson number:
\begin{equation}
N_{\text{av}}(t) = \text{Tr}\left[ \rho(t)|{N{\text{D}}\times N_{\text{D}}} \times ( a^{\dagger}a )|{N{\text{D}}\times N_{\text{D}}} \right]
\end{equation}
and the mutual information between spins and bosons:
\begin{equation}
I_{\text{spin-boson}} = S_{\text{spin}} + S_{\text{boson}} - S_{\text{spin-boson}},
\label{Eq:S}
\end{equation}
where $S_{\text{spin-boson}} = -\text{Tr}\left[\rho(t)\ln(\rho(t))\right]$,
$S_{\text{spin(boson)}} = -\text{Tr}\left[\rho_{\text{spin(boson)}}(t)\ln(\rho_{\text{spin(boson)}}(t))\right]$
are the corresponding entropies and
$\rho_{\text{spin}} (t)|_{(N+1) \times (N+1)} = \text{Tr}_{\text{boson}}\left[\rho(t)|_{N_{\text{D}}\times N_{\text{D}}} \right]$,
$\rho_{\text{boson}} (t)|_{(n_{\text{max}}+1) \times (n_{\text{max}}+1)} = \text{Tr}_{\text{spin}}\left[\rho(t)|_{N_{\text{D}}\times N_{\text{D}}} \right]$
is the corresponding reduced density matrix of
the spin(boson) obtained by tracing over the bosonic(spin) degrees
of freedom.
In Fig.~\ref{fig:qd_TM}, we consider low-energy initial states subjected to the Thue–Morse driving protocol. We present the time evolution of the average boson number [Fig.~\ref{fig:qd_TM}(a, b, e, f, i, j, m, n)] and the mutual information between spins and bosons [Fig.~\ref{fig:qd_TM}(c, d, g, h, k, l, o, p)]. These quantities are shown for a fixed intermediate driving frequency $\omega_{\text{d}}$ and various values of the coupling parameter $g_2$ [Fig.~\ref{fig:qd_TM}(a, c, e, g, i, k, m, o)]. In addition, we analyze the same observables for a fixed $(g_1, g_2)$ pair corresponding to the chaotic undriven model while varying the driving frequency $\omega_{\text{d}}$ [Fig.~\ref{fig:qd_TM}(b, d, f, h, j, l, n, p)].

The key observations in this case are given below.
\begin{enumerate}
\item Fig.~\ref{fig:qd_TM}(a-d) corresponds to $\kappa = 0$, and shows a
result similar to the closed quasiperiodic driven ADM.

\item For small values of $\kappa = 0.05$, we show that further heating is minimized, since both $I_{\text{spin-boson}}$, $N_{av}(t)$ are saturated less than the page value (Fig.~\ref{fig:qd_TM}(e-h)).

\item For increasing $\kappa$, the saturation value is further reduced. At the same time, for the non-ergodic phase, this saturation value is less than its ergodic counterpart (Fig.~\ref{fig:qd_TM}(i-p)). 
\item For the higher $\kappa$ regime, the saturation value and the corresponding saturation time are independent of $\omega_d$ (Fig.~\ref{fig:qd_TM}(n,p)).

\end{enumerate}
In summary, dissipation plays a constructive role in stabilizing the long-lived prethermal phase. In the next section, we also discuss the possible implications of our results for the rectification of an erroneous Floquet drive.

\section{Outlook}
\label{sec-v}
As a summary, we present a brief analysis of the existing phases in ADM in the presence of dissipation. The QPT in this case is modified by anisotropy  $(g_1 \neq g_2)$ and finite dissipation $(\kappa > 0)$. The ADM also shows ENET, as the vicinity of $g_1 = 0$, or $g_2 = 0$ shows the notion of non-ergodic phase, and the remaining part shows the ergodic behavior. In this work, we show that dissipation doesn't destroy the existence of ENET. We support our claim with the study of the Liouvillian ($\mathcal{L}$) of the atom+cavity. The Liouvillian gap ($\Delta$) shows two different scalings of ergodic and non-ergodic phases for changing the atom numbers. Similarly, the average participation ratio of the eigenvectors explicitly shows the notion of ENET.

Finally, we consider the driven ADM, where the periodic drive leads the system to the prethermal phase. A modification of the periodic Floquet drive (e.g., quasi-periodic drive) further introduces the heating process of the prethermal state. We find that the presence of dissipation is beneficial in this case. It is known that reservoir engineering protocol can be used for the ground state or targeted state preparation, as the unwanted excitation can be released through the leaking cavity to the environment. Motivated by this fact, we aim to investigate a similar situation here. Our result indicates that dissipation halts the excessive heating for the quasi-periodic case (Fig.~\ref{fig:qd_TM}(e-p)).  
For the experimental realization, we need to design a driving protocol that interpolates between the
two types of drives. $f(t) = a_1 \cos(\omega_d t) + a_2 \cos(\alpha \omega_d t)$. Here $\alpha$ is the irrational number. In the case of nuclear magnetic resonance, such a quasi-periodic drive is already implemented by providing two periodic delta kicks in the `y' direction, with $\alpha = (\sqrt{5} + 1)/2$ to observe discrete time quasi-crystals~\cite{he2025experimental}.
 
We also present possible future directions for our research. An extension of our analysis to TPT and ESQPT that incorporates dissipation is required. Similarly, the study of discrete time crystals in dissipative ADM under periodic driving would be of interest for understanding dynamical phases. Furthermore, the rapidly increasing dimensionality of the Liouvillian poses a challenge for simulating large atomic ensembles; addressing this issue will require the development of advanced numerical techniques in future work.

\section*{Acknowledgments}
We sincerely thank Auditya Sharma and Devendra Singh Bhakuni for fruitful discussions and constructive comments. We are also grateful to the High-Performance Computing (HPC) facility 
at IISER Bhopal. P. D. acknowledges IISERB for providing the PhD fellowship that supported the initial phase of this project. P. D. also acknowledges support from the European Research Council (ERC) through a postdoctoral fellowship (Project No. N1-0273). S. S. thanks the University Grants Commission for providing a PhD fellowship (Student ID: MAY 2018–528071), which supported the initial phase of this research. SS also acknowledges research support and funding from  BERLIN QUANTUM, an initiative endowed by the Innovation Promotion Fund of the city of Berlin, and the Deutsche Forschungsgemeinschaft (DFG, German Research Foundation) via the Research Unit FOR 5688 (Project No. 521530974). We also thank the International Centre for Theoretical Sciences (ICTS) for hosting the program Stability of Quantum Matter in and out of Equilibrium at Various Scales (ICTS/SQMVS2024/01), where this work and the associated discussions were initiated.

\appendix

    \section{Eigenvalue properties}
\label{app}
We provide a brief review of the distribution function of the eigenvalues
and level spacing ratios of a many-body quantum system, which
motivates us to study the nature of such a function for this model.
Based on the eigenvalue distribution of a Hermitian system, it can
be broadly classified into two categories: (i)
regular/integrable/non-ergodic and (ii) chaotic/ergodic. Berry and
Tabor hypothesized~\cite{berry1977level} that, for integrable
systems, the levels are uncorrelated, and hence the level spacing
distribution follows the one-dimensional Poisson distribution.
However, for correlated spectrum, the famous Wigner `surmise’~\cite{
bohr1969vol} is a powerful tool to investigate the spacing distribution.
In $1984$ Bohigas, Giannoni, and Schmit (BGS) hypothesized~\cite{bohigas1984characterization} that chaotic systems
would follow statistical behavior described by random matrix theory
appropriate to their symmetry classes. Following BGS, Grobe, Haake,
and Sommers (GHS)~\cite{grobe1988quantum} expanded the above spectral
contrast between integrability and chaos for Markovian open quantum
systems.
They found that in the integrable limit, the nearest neighbor spacing
distribution of radial distance $s$ for the complex eigenvalues of
Liouvillian agrees with the two-dimensional (2D) Poisson distribution.
Such a distribution is the same as Wigner’s conjecture for the spacing
distribution in the Gaussian orthogonal ensemble (GOE). On the other
hand, for a fully chaotic limit, the spacing distribution agrees with
the Ginibre ensemble ~\cite{ginibre1965statistical, grobe1988quantum,
markum1999non, haake1991quantum, hamazaki2020universality}.

Here, we study the eigenvalue properties of the Liouvillian
$\mathcal{L}$. Since complex eigenvalues are involved, we consider
generalized notions of the level spacing distribution and the consecutive 
level spacing ratio ~\cite{prasad2022dissipative} . By analyzing these 
quantities as a function of the coupling parameters, we try to sketch 
the phase diagram, which carries information on the non-ergodic to 
ergodic phase transition in the dissipative ADM. 

\begin{figure}[t]
\subfigure{\includegraphics[width=0.235\textwidth]{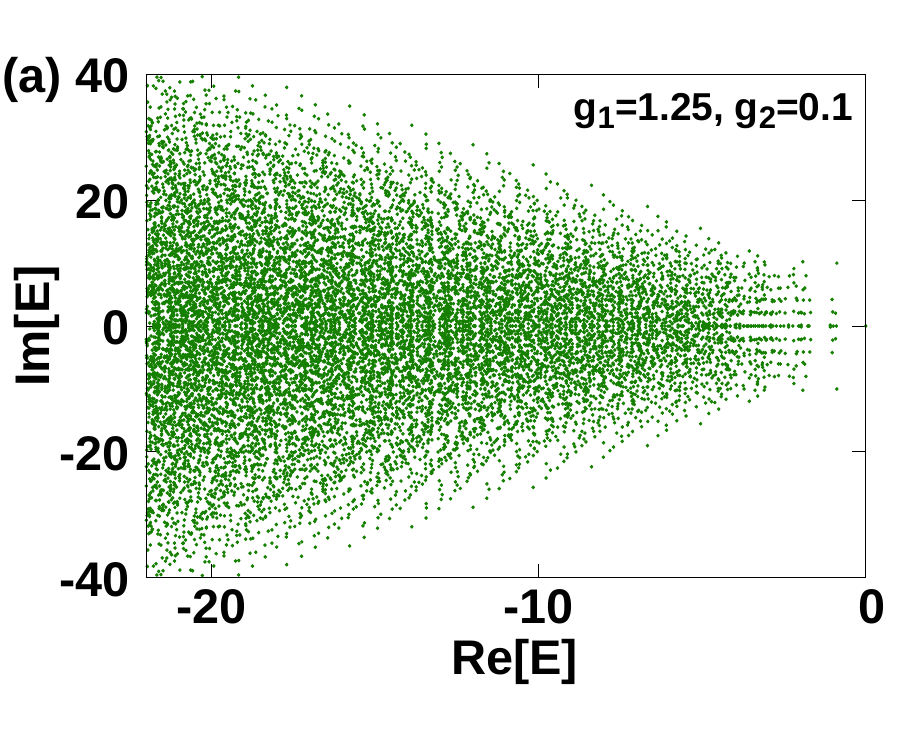}\label{fig:gs_energy_density}}
\subfigure{\includegraphics[width=0.235\textwidth]{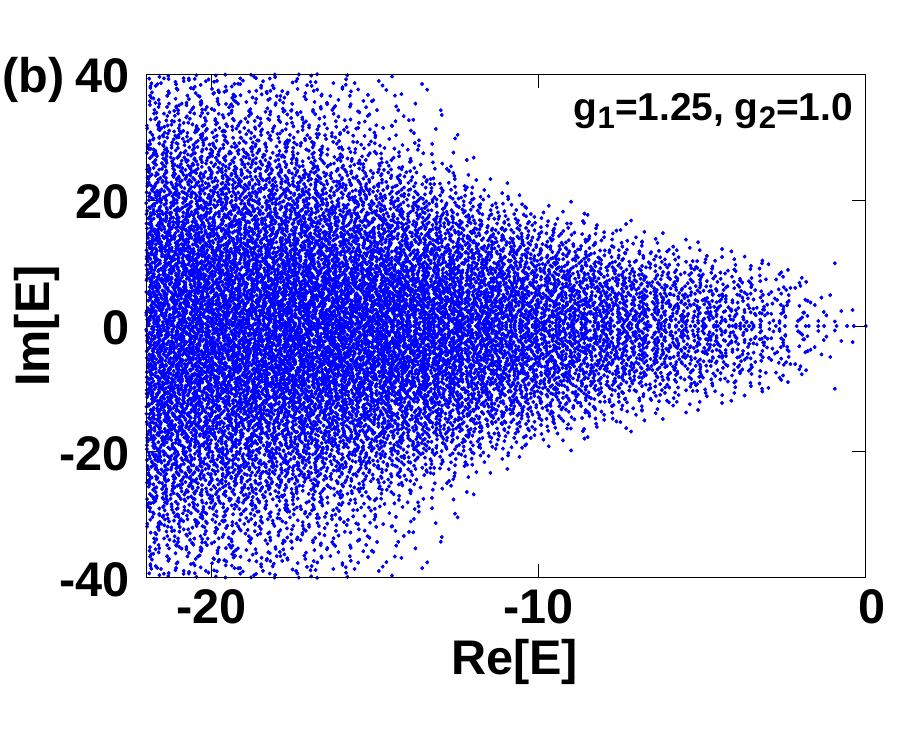}\label{fig:gs_energy_density}}
\caption{Scatter plot of the complex spectrum of the Liouvillian 
$\mathcal{L}$ of the dissipative anisotropic Dicke model for the atom 
number $N=10$ and $\omega=\omega_0=\kappa=1$, for (a) a point in the 
non-ergodic phase: $g_1=1.25,\ g_2=0.1$, and (b) a point in the ergodic 
phase: $g_1=1.25,\ g_2=1.0$. The bosonic cut-off: $n_{\text{max}} = 26$.  }
\label{fig:energy}
\end{figure}
 The eigenvalues of $\mathcal{L}$ lie in the nonpositive plane, which prevents the divergence of Eq.(\ref{eqn:lindblad}). The non-real eigenvalues possess complex conjugates, demonstrating the hermiticity of the density matrix. The zero eigenvalue of $\mathcal{L}$ corresponds to a steady state. 
The real part of the complex eigenvalues corresponds to the loss or decay of the initial density matrix \cite{minganti2018spectral}.

In Fig.~\ref{fig:energy}, we show the scatter plot of the complex 
spectrum of the Liouvillian $\mathcal{L}$ for two different set of $(g_1, g_2)$ pairs: $g_1=1.25$ and 
(a) $g_2 = 0.1$, (b) $g_2 = 1.0$ corresponding to the non-ergodic and 
ergodic regimes respectively.     
They clearly show the different natures of the spectrum in the regular and chaotic regimes, respectively. For a clearer understanding of the two different phases, one should study the level spacing statistics. Additionally, we would like to discuss briefly how to calculate the level spacing distribution for a non-Hermitian system.

We first
carry out an unfolding of the spectrum to remove the system-specific
aspects of the level spacing statistics.
The complex level spacing distribution $P(s)$ is then produced by
using the unfolded spectrum to create a histogram of the Euclidean
distance $s$ between nearest-neighbor eigenvalues in the complex plane.
For the unfolding process, we mainly follow the references~\cite{
haake1991quantum, akemann2019universal, hamazaki2020universality}.

The Euclidean distance of the nearest neighbor (NN) of each eigenvalue 
of the Liouvillian is given by, $s_i \equiv |E_i - E_i^{\text{NN}} |$. 
After rescaling, the analytical expression is given by:
 \begin{equation}
 s_i \rightarrow s_i^{\prime} = s_i \frac{\sqrt{\rho_{\text{av}}(E_i)}}{\bar{s}}.
 \label{eqn:s_prime}
\end{equation}      

Here $\rho_{\text{av}}(E_i)$ is the local average density, which
can be approximated in the following way:
\begin{equation}
\rho_{\text{av}}(E_i) = \frac{1}{2\pi\sigma^2 N}\sum_{i=1}^{N} \exp\left( - \frac{|E-E_i|^2}{2\sigma^2} \right).
\label{eqn:rho_av}
\end{equation}
Following the earlier references~\cite{haake1991quantum,
akemann2019universal, hamazaki2020universality}, we choose:
$\sigma = 4.5\times \bar{s}$, where $\bar{s} = \frac{1}{N}\sum_{i=1}
^{N} s_i$. In Eq~\ref{eqn:s_prime} $\bar{s}$ is chosen such a way
that: $\frac{1}{N}\sum_{i=1}^{N} s_i^{\prime} = 1$. Finally, we
will study the level statistics of the NN distance $s_i^{\prime}$.

\begin{figure}[t]
  \subfigure{\includegraphics[width=0.235\textwidth]{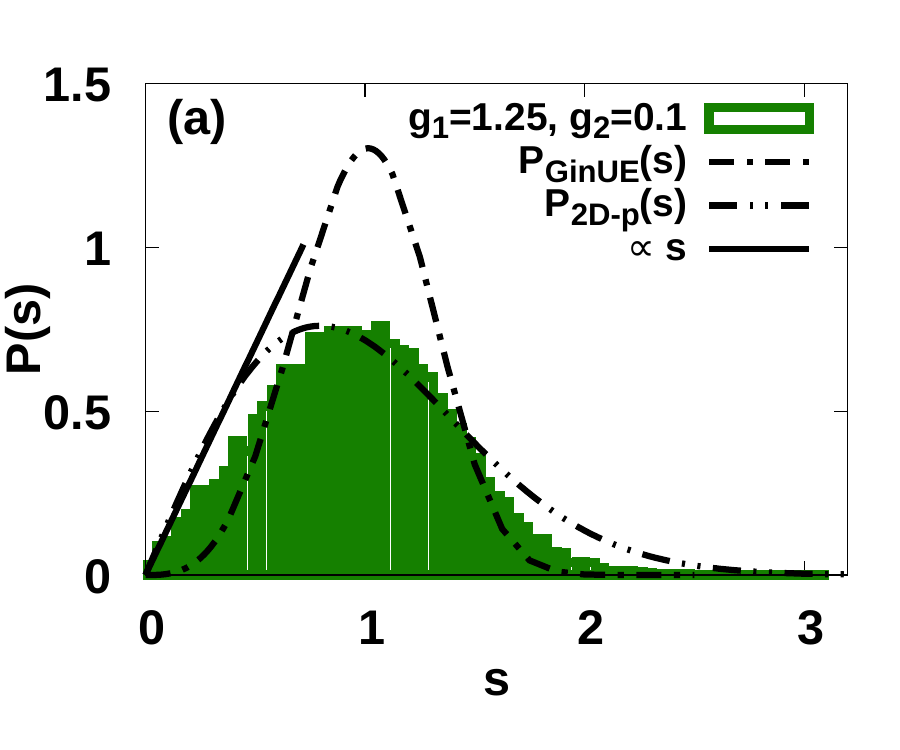}\label{fig:gs_energy_density}}
  \subfigure{\includegraphics[width=0.235\textwidth]{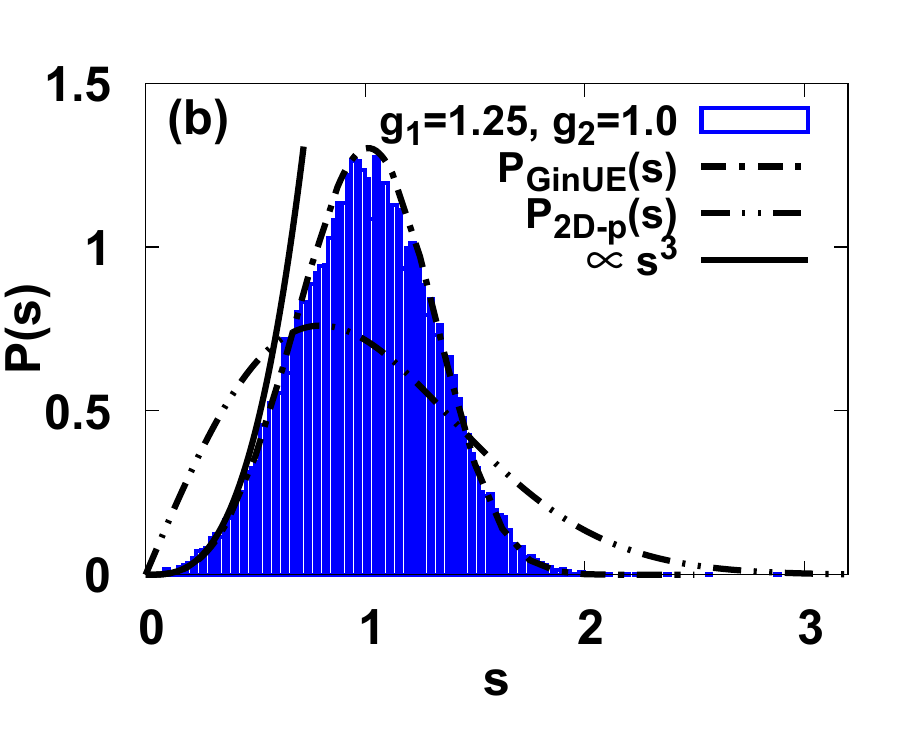}\label{fig:gs_energy_density}}
  \caption{Level-spacing distribution of the complex spectrum of the 
  Liouvillian $\mathcal{L}$ in (a) the non-ergodic phase with $g_1=1.25,\ 
  g_2=0.1$ and (b) the ergodic phase with $g_1=1.25,\ g_2=1.0$. We 
  find remarkable agreement with the 2D Poisson distribution 
  $P_{\text{2D-P}}(s)$ in Eq. and that of the GinUE RMT prediction 
  $P_{\text{GinUE}}(s)$. The atom number $N=10$, bosonic 
  cut-off: $n_{\text{max}} = 26$. The other parameters are: 
  $\omega=\omega_0=1$, $\kappa=1$. }
  \label{fig:spacing_distribution}
\end{figure}

The analytical form of $2D$ Poisson distribution, which corresponds to 
the spacing distribution for independent complex random values is given by:
\begin{equation}
P_{2D-P}(s) = \frac{\pi}{2} s \exp\left( - \frac{\pi s^2}{4} \right).
\label{eqn:poi_2D}
\end{equation}  
On the other hand, for the fully chaotic (or ergodic) limit, the spacing 
distribution agrees with the corresponding distribution of the Ginibre 
ensemble of complex Gaussian non-Hermitian random matrices, 
given by:
\begin{equation}
P_{\text{GinUE}}(s) = \bar{s}\bar{P}_{\text{GinUE}}(\bar{s}s),
\label{eqn:gini}
\end{equation}
with
\begin{equation}
\bar{P}_{\text{GinUE}}(s) = \sum_{j=1}^{\infty}\frac{2 s^{2j + 1}\exp(-s^2)}{\Gamma(1+j, s^2)}\Pi_{k=1}^{\infty}\frac{\Gamma(1+k, s^2)}{k!}
\end{equation}
\begin{figure}[t]
  \subfigure{\includegraphics[width=0.235\textwidth]{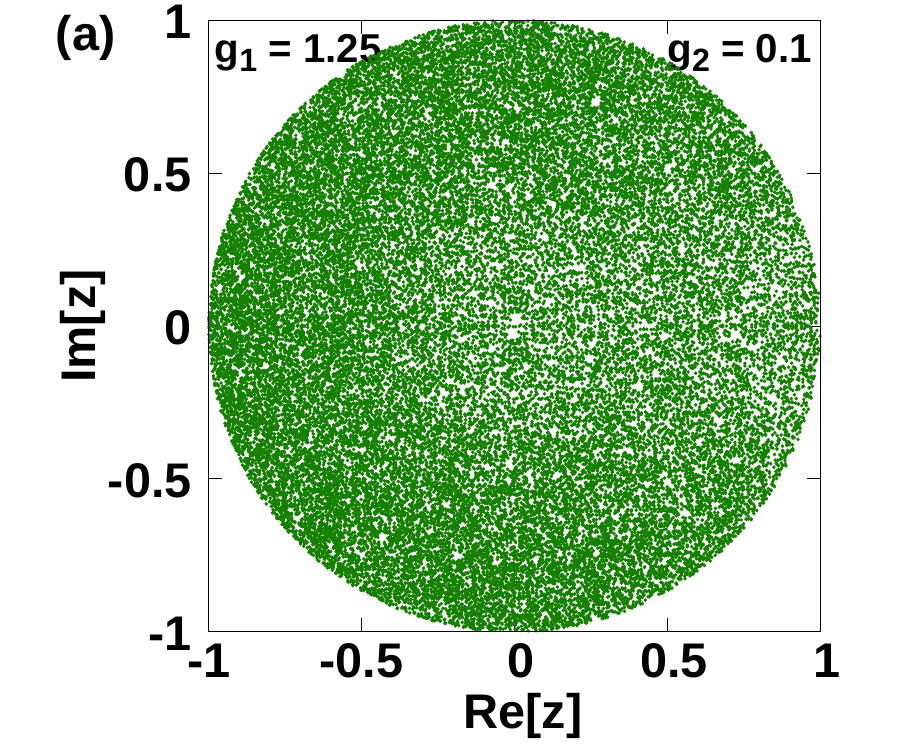}\label{fig:gs_energy_density}}
  \subfigure{\includegraphics[width=0.235\textwidth]{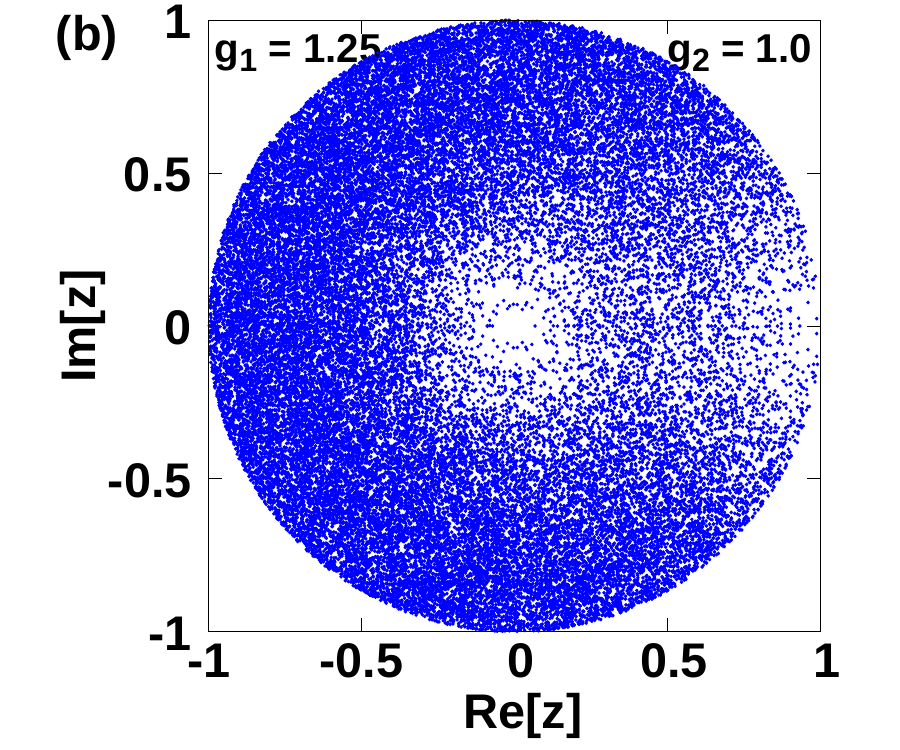}\label{fig:gs_energy_density}}
  \caption{Scatter plot of the complex level-spacing ratio $z$ introduced 
  in Eq.~\ref{eqn:spacing_ratio} for the atom number $N=10$, for (a) a point in the non-ergodic phase: 
  $g_1=1.25,\ g_2=0.1$, which shows the isotropic behavior and (b) a point in 
  the ergodic pgase: $g_1=1.25,\ g_2=1.0$, which is not isotropic. The bosonic 
  cut-off: $n_{\text{max}} = 26$. The other parameters are: 
  $\omega=\omega_0=1$, $\kappa=1$. }
  \label{fig:r_scatter}
\end{figure}
and $\bar{s} = \int_{0}^{\infty} s \bar{P}_{\text{GinUE}}(s)\hspace{1mm} ds$, 
$\Gamma(1+k, s^2) = \int_{s^2}^{\infty} t^k e^{-t} dt$, is the incomplete 
Gamma function. 

We study the level spacing statistics separately for these two pair 
points in different phases (see Fig.~\ref{fig:spacing_distribution}).
Fig.~\ref{fig:spacing_distribution} displays the striking agreement 
between the distributions calculated from the spectrum of $\mathcal{L}$ 
and the $2D$ Poisson distribution in the integrable or regular or the 
non-ergodic regime and the GinUE forecast in the chaotic or ergodic 
regime, respectively. 
In the ergodic phase $P(s) \approx s^3$, which indicates the chaotic 
behavior of the complex eigenvalues as they repel each other, and this 
is consistent with Ginibre distribution in Eq.~\ref{eqn:gini}. On 
the other hand, in the non-ergodic phase, we have $P(s) \approx s$ 
indicating the integrable nature of this regime and consistent with 
$2D$ Poisson distribution in the complex plane. 

\begin{figure}[t]
  \includegraphics[width=0.35\textwidth]{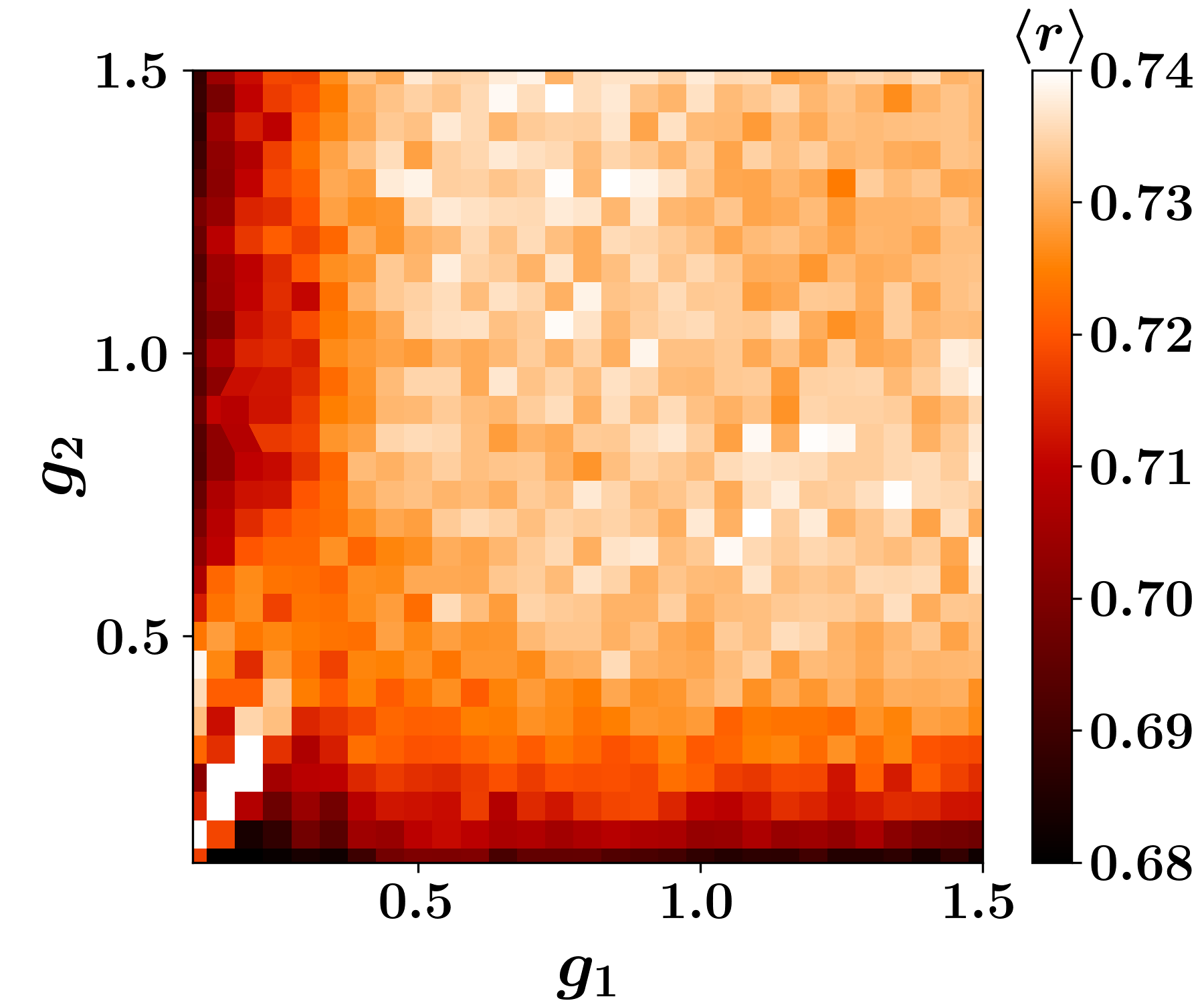}
  \caption{Average level spacing ratio, $\langle r\rangle$ of the dissipative 
  anisotropic Dicke model. It shows a transition from the non-ergodic phase 
  (where the system follow the $2D$ Poison statics and $\langle r\rangle
  \approx 0.69$) to the ergodic phase (where the system behaves as the Ginibre 
  ensemble and $\langle r\rangle\approx 0.74$). Here the atom number $N=8$, 
  the bosonic cut-off $n_{\text{max}}=20$. The other parameters are: 
  $\omega=\omega_0=1$, $\kappa=1$.}
  \label{fig:rav_OADM}
\end{figure}

    \subsection{Consecutive level spacing ratio } 
    The complex level spacing 
    ratio~\cite{sa2020complex, prasad2022dissipative} can be defined as:
    \begin{equation}
    z_i = r_i e^{\text{i}\theta_i}=\frac{E_i^{\text{NN}}-E_i}{E_i^{\text{NNN}}-E_i},
    \label{eqn:spacing_ratio}
    \end{equation}
    where the superscripts NN (NNN) stand for the nearest (next nearest) 
    neighbor. It captures the information about the next closest neighbor that is 
    overlooked in the level spacing statistics. 
    The advantage of the calculation is that we don't require the 
    unfolding procedure. 
    In Fig.~\ref{fig:r_scatter}, we show the scatter plot of $z$ in 
    the non-ergodic ($g_1 = 1.25,\ g_2 = 0.1$) and ergodic phases 
    ($g_1 = 1.25,\ g_2 = 1.0$) separately. For the non-ergodic case, the 
    scatter plot is symmetric and isotropic, whereas for the ergodic case, 
    it is anisotropic, which is another signature for the connection to 
    RMT~\cite{sa2020complex}. In Fig.~\ref{fig:rav_OADM} we show the 
    absolute value of the average level spacing ratio for this model. 
    The figure shows a non-ergodic phase to ergodic phase transition 
    similar to the closed ADM~\cite{das2023phase}. 
    Here in the non-ergodic phase, $\langle r\rangle\approx 0.69$, 
    (dark color) which is equivalent to the system following the 2D 
    Poisson distribution, whereas in the ergodic phase, $\langle r\rangle\approx 0.74$ (light color), and it corresponds to the Ginibre unitary 
    ensemble.

\section{Dependence of the exponent $z$ by varying $g_1$ and $g_2$.}
\label{app3}
We provide a contour plot of the exponent $z$ by varying $g_1$ and $g_2$ for $\kappa = 0.1$. 
\begin{figure}[h]
\subfigure{\includegraphics[width=0.35\textwidth]{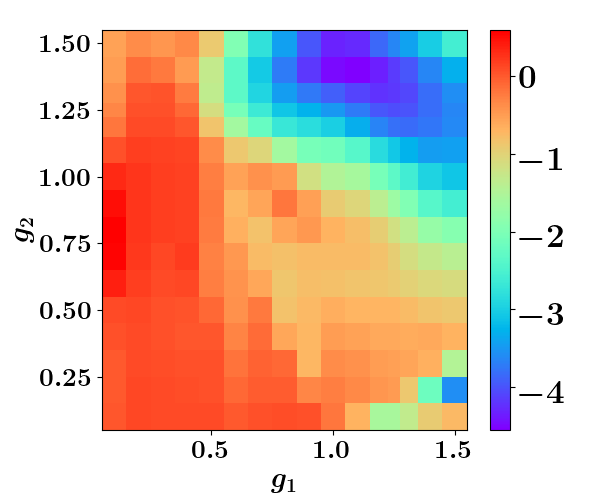}}
  \caption{The exponent $z$ (extracted from the scaling of Liouvillian gap $\Delta$ as a function of atom number) as a function of $g_1$ and $g_2$ for  $\kappa=0.1$ and $n_{\text{max}}=26$. Here, the atom number $N$ is changing from $4:10$. For varying $g_1$ and $g_2$, the system shows a response which has similarities to ENET. For example, in the vicinity of $g_1 =0$ and $g_2 =0$ (non-ergodic phase), the value of $z$ (dark-orange portion) is different from the remaining part (ergodic phase).}
  \label{fig:L_gap_cont}
\end{figure}
Our results successfully capture the key features of the system, even though the extreme point ($g_1 =g_2$), which occurs farthest from the integrability, is not yet clearly visible. Nevertheless, the dependence of $z$ on $g_1$ and $g_2$ shows qualitative similarities to ENET. To achieve a clearer understanding, simulations with a larger number of atoms and a higher bosonic cutoff will be required.

\section{Average participation ratio for different $\kappa$ }
 \label{app2}
\begin{figure*}
  \subfigure[]
  {\includegraphics[width=0.325\textwidth]{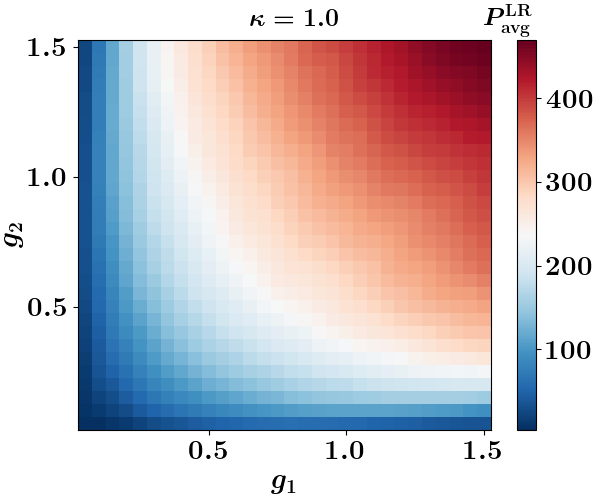}}
  \subfigure[]
  {\includegraphics[width=0.325\textwidth]{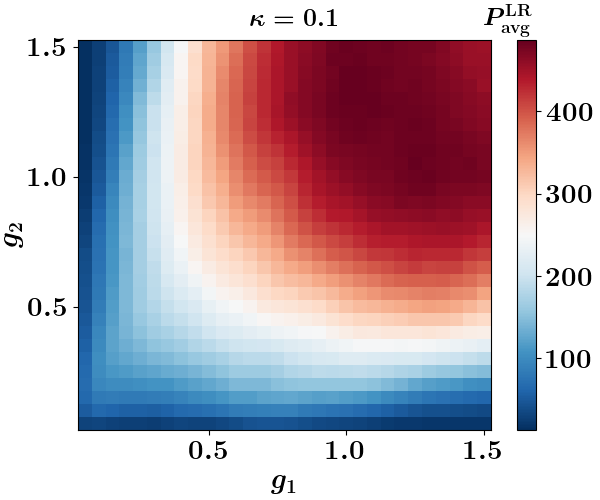}}
  \subfigure[]
  {\includegraphics[width=0.325\textwidth]{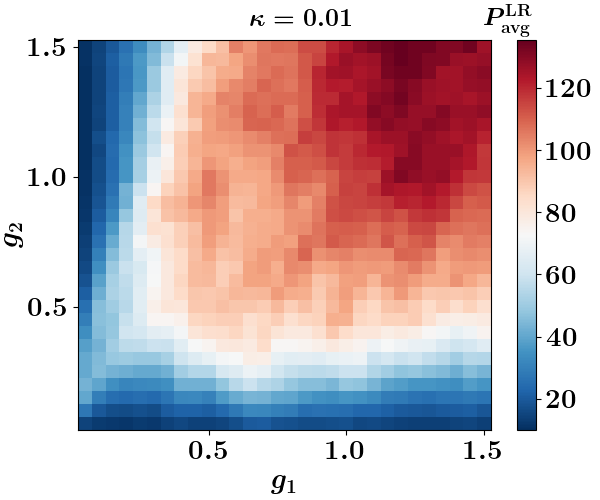}}
  \caption{
  Average of participation ratio ($P^{\rm LR}_{\text{avg}}$), using biorthonormal basis considering both left 
    and right eigenvectors of the Liouvillian for dissipative ADM  as a 
    function of the rotating and counter-rotating coupling parameters $g_1$ 
    and $g_2$. The atom number: $N=4$, the bosonic cut-off: $n_{\text{max}}
    =10$. For this figures $\omega_0 = \omega=1,\ \kappa=1.0, \, 0.1, \, 0.01$ respectively.  
  }
  \label{fig:P_avg_LR}
\end{figure*}

Here we provide the contour plots of $P^{\rm LR}_{\rm avg}$ for different $\kappa$ by varying $g_1$ and $g_2$ in Fig.~\ref{fig:P_avg_LR}. We observe that the study of $P^{\rm LR}_{\rm avg}$ successfully captures the ENET, and for changing $\kappa$, the qualitative features remain unchanged.

\twocolumngrid
\newpage
\bibliography{ref}	

\begin{thebibliography}{83}%
\makeatletter
\providecommand \@ifxundefined [1]{%
 \@ifx{#1\undefined}
}%
\providecommand \@ifnum [1]{%
 \ifnum #1\expandafter \@firstoftwo
 \else \expandafter \@secondoftwo
 \fi
}%
\providecommand \@ifx [1]{%
 \ifx #1\expandafter \@firstoftwo
 \else \expandafter \@secondoftwo
 \fi
}%
\providecommand \natexlab [1]{#1}%
\providecommand \enquote  [1]{``#1''}%
\providecommand \bibnamefont  [1]{#1}%
\providecommand \bibfnamefont [1]{#1}%
\providecommand \citenamefont [1]{#1}%
\providecommand \href@noop [0]{\@secondoftwo}%
\providecommand \href [0]{\begingroup \@sanitize@url \@href}%
\providecommand \@href[1]{\@@startlink{#1}\@@href}%
\providecommand \@@href[1]{\endgroup#1\@@endlink}%
\providecommand \@sanitize@url [0]{\catcode `\\12\catcode `\$12\catcode
  `\&12\catcode `\#12\catcode `\^12\catcode `\_12\catcode `\%12\relax}%
\providecommand \@@startlink[1]{}%
\providecommand \@@endlink[0]{}%
\providecommand \url  [0]{\begingroup\@sanitize@url \@url }%
\providecommand \@url [1]{\endgroup\@href {#1}{\urlprefix }}%
\providecommand \urlprefix  [0]{URL }%
\providecommand \Eprint [0]{\href }%
\providecommand \doibase [0]{http://dx.doi.org/}%
\providecommand \selectlanguage [0]{\@gobble}%
\providecommand \bibinfo  [0]{\@secondoftwo}%
\providecommand \bibfield  [0]{\@secondoftwo}%
\providecommand \translation [1]{[#1]}%
\providecommand \BibitemOpen [0]{}%
\providecommand \bibitemStop [0]{}%
\providecommand \bibitemNoStop [0]{.\EOS\space}%
\providecommand \EOS [0]{\spacefactor3000\relax}%
\providecommand \BibitemShut  [1]{\csname bibitem#1\endcsname}%
\let\auto@bib@innerbib\@empty
\bibitem [{\citenamefont {Cohen-Tannoudji}\ \emph {et~al.}(1998)\citenamefont
  {Cohen-Tannoudji}, \citenamefont {Dupont-Roc},\ and\ \citenamefont
  {Grynberg}}]{cohen-tannoudji_atom-photon_1998}%
  \BibitemOpen
  \bibfield  {author} {\bibinfo {author} {\bibfnamefont {C.}~\bibnamefont
  {Cohen-Tannoudji}}, \bibinfo {author} {\bibfnamefont {J.}~\bibnamefont
  {Dupont-Roc}}, \ and\ \bibinfo {author} {\bibfnamefont {G.}~\bibnamefont
  {Grynberg}},\ }\href {https://books.google.de/books?id=hNWbEAAAQBAJ} {\emph
  {\bibinfo {title} {Atom-{Photon} {Interactions}: {Basic} {Processes} and
  {Applications}}}}\ (\bibinfo  {publisher} {Wiley},\ \bibinfo {year}
  {1998})\BibitemShut {NoStop}%
\bibitem [{\citenamefont {L{\'e}onard}\ \emph {et~al.}(2017)\citenamefont
  {L{\'e}onard}, \citenamefont {Morales}, \citenamefont {Zupancic},
  \citenamefont {Donner},\ and\ \citenamefont
  {Esslinger}}]{leonard2017monitoring}%
  \BibitemOpen
  \bibfield  {author} {\bibinfo {author} {\bibfnamefont {J.}~\bibnamefont
  {L{\'e}onard}}, \bibinfo {author} {\bibfnamefont {A.}~\bibnamefont
  {Morales}}, \bibinfo {author} {\bibfnamefont {P.}~\bibnamefont {Zupancic}},
  \bibinfo {author} {\bibfnamefont {T.}~\bibnamefont {Donner}}, \ and\ \bibinfo
  {author} {\bibfnamefont {T.}~\bibnamefont {Esslinger}},\ }\href@noop {}
  {\bibfield  {journal} {\bibinfo  {journal} {Science}\ }\textbf {\bibinfo
  {volume} {358}},\ \bibinfo {pages} {1415} (\bibinfo {year}
  {2017})}\BibitemShut {NoStop}%
\bibitem [{\citenamefont {Gutzler}\ \emph {et~al.}(2021)\citenamefont
  {Gutzler}, \citenamefont {Garg}, \citenamefont {Ast}, \citenamefont
  {Kuhnke},\ and\ \citenamefont {Kern}}]{gutzler_lightmatter_2021}%
  \BibitemOpen
  \bibfield  {author} {\bibinfo {author} {\bibfnamefont {R.}~\bibnamefont
  {Gutzler}}, \bibinfo {author} {\bibfnamefont {M.}~\bibnamefont {Garg}},
  \bibinfo {author} {\bibfnamefont {C.~R.}\ \bibnamefont {Ast}}, \bibinfo
  {author} {\bibfnamefont {K.}~\bibnamefont {Kuhnke}}, \ and\ \bibinfo {author}
  {\bibfnamefont {K.}~\bibnamefont {Kern}},\ }\href {\doibase
  10.1038/s42254-021-00306-5} {\bibfield  {journal} {\bibinfo  {journal}
  {Nature Reviews Physics}\ }\textbf {\bibinfo {volume} {3}},\ \bibinfo {pages}
  {441} (\bibinfo {year} {2021})},\ \bibinfo {note} {publisher: Nature
  Publishing Group}\BibitemShut {NoStop}%
\bibitem [{\citenamefont {Chang}\ \emph {et~al.}(2018)\citenamefont {Chang},
  \citenamefont {Douglas}, \citenamefont {Gonz\'alez-Tudela}, \citenamefont
  {Hung},\ and\ \citenamefont {Kimble}}]{RevModPhys.90.031002}%
  \BibitemOpen
  \bibfield  {author} {\bibinfo {author} {\bibfnamefont {D.~E.}\ \bibnamefont
  {Chang}}, \bibinfo {author} {\bibfnamefont {J.~S.}\ \bibnamefont {Douglas}},
  \bibinfo {author} {\bibfnamefont {A.}~\bibnamefont {Gonz\'alez-Tudela}},
  \bibinfo {author} {\bibfnamefont {C.-L.}\ \bibnamefont {Hung}}, \ and\
  \bibinfo {author} {\bibfnamefont {H.~J.}\ \bibnamefont {Kimble}},\ }\href
  {\doibase 10.1103/RevModPhys.90.031002} {\bibfield  {journal} {\bibinfo
  {journal} {Rev. Mod. Phys.}\ }\textbf {\bibinfo {volume} {90}},\ \bibinfo
  {pages} {031002} (\bibinfo {year} {2018})}\BibitemShut {NoStop}%
\bibitem [{\citenamefont {Walther}\ \emph {et~al.}(2006)\citenamefont
  {Walther}, \citenamefont {Varcoe}, \citenamefont {Englert},\ and\
  \citenamefont {Becker}}]{walther_cavity_2006}%
  \BibitemOpen
  \bibfield  {author} {\bibinfo {author} {\bibfnamefont {H.}~\bibnamefont
  {Walther}}, \bibinfo {author} {\bibfnamefont {B.~T.~H.}\ \bibnamefont
  {Varcoe}}, \bibinfo {author} {\bibfnamefont {B.-G.}\ \bibnamefont {Englert}},
  \ and\ \bibinfo {author} {\bibfnamefont {T.}~\bibnamefont {Becker}},\ }\href
  {\doibase 10.1088/0034-4885/69/5/R02} {\bibfield  {journal} {\bibinfo
  {journal} {Reports on Progress in Physics}\ }\textbf {\bibinfo {volume}
  {69}},\ \bibinfo {pages} {1325} (\bibinfo {year} {2006})}\BibitemShut
  {NoStop}%
\bibitem [{\citenamefont {Reiserer}\ and\ \citenamefont
  {Rempe}(2015)}]{RevModPhys.87.1379}%
  \BibitemOpen
  \bibfield  {author} {\bibinfo {author} {\bibfnamefont {A.}~\bibnamefont
  {Reiserer}}\ and\ \bibinfo {author} {\bibfnamefont {G.}~\bibnamefont
  {Rempe}},\ }\href {\doibase 10.1103/RevModPhys.87.1379} {\bibfield  {journal}
  {\bibinfo  {journal} {Rev. Mod. Phys.}\ }\textbf {\bibinfo {volume} {87}},\
  \bibinfo {pages} {1379} (\bibinfo {year} {2015})}\BibitemShut {NoStop}%
\bibitem [{\citenamefont {Vaidya}\ \emph {et~al.}(2018)\citenamefont {Vaidya},
  \citenamefont {Guo}, \citenamefont {Kroeze}, \citenamefont {Ballantine},
  \citenamefont {Koll\'ar}, \citenamefont {Keeling},\ and\ \citenamefont
  {Lev}}]{PhysRevX.8.011002}%
  \BibitemOpen
  \bibfield  {author} {\bibinfo {author} {\bibfnamefont {V.~D.}\ \bibnamefont
  {Vaidya}}, \bibinfo {author} {\bibfnamefont {Y.}~\bibnamefont {Guo}},
  \bibinfo {author} {\bibfnamefont {R.~M.}\ \bibnamefont {Kroeze}}, \bibinfo
  {author} {\bibfnamefont {K.~E.}\ \bibnamefont {Ballantine}}, \bibinfo
  {author} {\bibfnamefont {A.~J.}\ \bibnamefont {Koll\'ar}}, \bibinfo {author}
  {\bibfnamefont {J.}~\bibnamefont {Keeling}}, \ and\ \bibinfo {author}
  {\bibfnamefont {B.~L.}\ \bibnamefont {Lev}},\ }\href {\doibase
  10.1103/PhysRevX.8.011002} {\bibfield  {journal} {\bibinfo  {journal} {Phys.
  Rev. X}\ }\textbf {\bibinfo {volume} {8}},\ \bibinfo {pages} {011002}
  (\bibinfo {year} {2018})}\BibitemShut {NoStop}%
\bibitem [{\citenamefont {Dimer}\ \emph
  {et~al.}(2007{\natexlab{a}})\citenamefont {Dimer}, \citenamefont {Estienne},
  \citenamefont {Parkins},\ and\ \citenamefont
  {Carmichael}}]{dimer2007proposed}%
  \BibitemOpen
  \bibfield  {author} {\bibinfo {author} {\bibfnamefont {F.}~\bibnamefont
  {Dimer}}, \bibinfo {author} {\bibfnamefont {B.}~\bibnamefont {Estienne}},
  \bibinfo {author} {\bibfnamefont {A.~S.}\ \bibnamefont {Parkins}}, \ and\
  \bibinfo {author} {\bibfnamefont {H.~J.}\ \bibnamefont {Carmichael}},\ }\href
  {\doibase 10.1103/PhysRevA.75.013804} {\bibfield  {journal} {\bibinfo
  {journal} {Phys. Rev. A}\ }\textbf {\bibinfo {volume} {75}},\ \bibinfo
  {pages} {013804} (\bibinfo {year} {2007}{\natexlab{a}})}\BibitemShut
  {NoStop}%
\bibitem [{\citenamefont {Garraway}(2011)}]{garraway_dicke_2011}%
  \BibitemOpen
  \bibfield  {author} {\bibinfo {author} {\bibfnamefont {B.~M.}\ \bibnamefont
  {Garraway}},\ }\href {\doibase 10.1098/rsta.2010.0333} {\bibfield  {journal}
  {\bibinfo  {journal} {Philosophical Transactions of the Royal Society A:
  Mathematical, Physical and Engineering Sciences}\ }\textbf {\bibinfo {volume}
  {369}},\ \bibinfo {pages} {1137} (\bibinfo {year} {2011})},\ \bibinfo {note}
  {publisher: Royal Society}\BibitemShut {NoStop}%
\bibitem [{\citenamefont {Kirton}\ \emph {et~al.}(2019)\citenamefont {Kirton},
  \citenamefont {Roses}, \citenamefont {Keeling},\ and\ \citenamefont
  {Dalla~Torre}}]{kirton2019introduction}%
  \BibitemOpen
  \bibfield  {author} {\bibinfo {author} {\bibfnamefont {P.}~\bibnamefont
  {Kirton}}, \bibinfo {author} {\bibfnamefont {M.~M.}\ \bibnamefont {Roses}},
  \bibinfo {author} {\bibfnamefont {J.}~\bibnamefont {Keeling}}, \ and\
  \bibinfo {author} {\bibfnamefont {E.~G.}\ \bibnamefont {Dalla~Torre}},\
  }\href {\doibase https://doi.org/10.1002/qute.201800043} {\bibfield
  {journal} {\bibinfo  {journal} {Advanced Quantum Technologies}\ }\textbf
  {\bibinfo {volume} {2}},\ \bibinfo {pages} {1800043} (\bibinfo {year}
  {2019})}\BibitemShut {NoStop}%
\bibitem [{\citenamefont {Villaseñor}\ \emph {et~al.}(2024)\citenamefont
  {Villaseñor}, \citenamefont {Pilatowsky-Cameo}, \citenamefont
  {Chávez-Carlos}, \citenamefont {Bastarrachea-Magnani}, \citenamefont
  {Lerma-Hernández}, \citenamefont {Santos},\ and\ \citenamefont
  {Hirsch}}]{villaseñor2024}%
  \BibitemOpen
  \bibfield  {author} {\bibinfo {author} {\bibfnamefont {D.}~\bibnamefont
  {Villaseñor}}, \bibinfo {author} {\bibfnamefont {S.}~\bibnamefont
  {Pilatowsky-Cameo}}, \bibinfo {author} {\bibfnamefont {J.}~\bibnamefont
  {Chávez-Carlos}}, \bibinfo {author} {\bibfnamefont {M.~A.}\ \bibnamefont
  {Bastarrachea-Magnani}}, \bibinfo {author} {\bibfnamefont {S.}~\bibnamefont
  {Lerma-Hernández}}, \bibinfo {author} {\bibfnamefont {L.~F.}\ \bibnamefont
  {Santos}}, \ and\ \bibinfo {author} {\bibfnamefont {J.~G.}\ \bibnamefont
  {Hirsch}},\ }\href {https://arxiv.org/abs/2405.20381} {\enquote {\bibinfo
  {title} {Classical and quantum properties of the spin-boson dicke model:
  Chaos, localization, and scarring},}\ } (\bibinfo {year} {2024}),\ \Eprint
  {http://arxiv.org/abs/2405.20381} {arXiv:2405.20381 [quant-ph]} \BibitemShut
  {NoStop}%
\bibitem [{\citenamefont {Dicke}(1954)}]{dicke1954coherence}%
  \BibitemOpen
  \bibfield  {author} {\bibinfo {author} {\bibfnamefont {R.~H.}\ \bibnamefont
  {Dicke}},\ }\href@noop {} {\bibfield  {journal} {\bibinfo  {journal}
  {Physical review}\ }\textbf {\bibinfo {volume} {93}},\ \bibinfo {pages} {99}
  (\bibinfo {year} {1954})}\BibitemShut {NoStop}%
\bibitem [{\citenamefont {Lambert}\ \emph {et~al.}(2004)\citenamefont
  {Lambert}, \citenamefont {Emary},\ and\ \citenamefont
  {Brandes}}]{lambert2004entanglement}%
  \BibitemOpen
  \bibfield  {author} {\bibinfo {author} {\bibfnamefont {N.}~\bibnamefont
  {Lambert}}, \bibinfo {author} {\bibfnamefont {C.}~\bibnamefont {Emary}}, \
  and\ \bibinfo {author} {\bibfnamefont {T.}~\bibnamefont {Brandes}},\ }\href
  {\doibase 10.1103/PhysRevLett.92.073602} {\bibfield  {journal} {\bibinfo
  {journal} {Phys. Rev. Lett.}\ }\textbf {\bibinfo {volume} {92}},\ \bibinfo
  {pages} {073602} (\bibinfo {year} {2004})}\BibitemShut {NoStop}%
\bibitem [{\citenamefont {Emary}\ and\ \citenamefont
  {Brandes}(2003{\natexlab{a}})}]{emary2003chaos}%
  \BibitemOpen
  \bibfield  {author} {\bibinfo {author} {\bibfnamefont {C.}~\bibnamefont
  {Emary}}\ and\ \bibinfo {author} {\bibfnamefont {T.}~\bibnamefont
  {Brandes}},\ }\href {\doibase 10.1103/PhysRevE.67.066203} {\bibfield
  {journal} {\bibinfo  {journal} {Phys. Rev. E}\ }\textbf {\bibinfo {volume}
  {67}},\ \bibinfo {pages} {066203} (\bibinfo {year}
  {2003}{\natexlab{a}})}\BibitemShut {NoStop}%
\bibitem [{\citenamefont {Emary}\ and\ \citenamefont
  {Brandes}(2003{\natexlab{b}})}]{emary2003quantum}%
  \BibitemOpen
  \bibfield  {author} {\bibinfo {author} {\bibfnamefont {C.}~\bibnamefont
  {Emary}}\ and\ \bibinfo {author} {\bibfnamefont {T.}~\bibnamefont
  {Brandes}},\ }\href {\doibase 10.1103/PhysRevLett.90.044101} {\bibfield
  {journal} {\bibinfo  {journal} {Phys. Rev. Lett.}\ }\textbf {\bibinfo
  {volume} {90}},\ \bibinfo {pages} {044101} (\bibinfo {year}
  {2003}{\natexlab{b}})}\BibitemShut {NoStop}%
\bibitem [{\citenamefont {Ch\'avez-Carlos}\ \emph {et~al.}(2016)\citenamefont
  {Ch\'avez-Carlos}, \citenamefont {Bastarrachea-Magnani}, \citenamefont
  {Lerma-Hern\'andez},\ and\ \citenamefont {Hirsch}}]{chavez2016classical}%
  \BibitemOpen
  \bibfield  {author} {\bibinfo {author} {\bibfnamefont {J.}~\bibnamefont
  {Ch\'avez-Carlos}}, \bibinfo {author} {\bibfnamefont {M.~A.}\ \bibnamefont
  {Bastarrachea-Magnani}}, \bibinfo {author} {\bibfnamefont {S.}~\bibnamefont
  {Lerma-Hern\'andez}}, \ and\ \bibinfo {author} {\bibfnamefont {J.~G.}\
  \bibnamefont {Hirsch}},\ }\href {\doibase 10.1103/PhysRevE.94.022209}
  {\bibfield  {journal} {\bibinfo  {journal} {Phys. Rev. E}\ }\textbf {\bibinfo
  {volume} {94}},\ \bibinfo {pages} {022209} (\bibinfo {year}
  {2016})}\BibitemShut {NoStop}%
\bibitem [{\citenamefont {Kirton}\ and\ \citenamefont
  {Keeling}(2018)}]{kirton2018superradiant}%
  \BibitemOpen
  \bibfield  {author} {\bibinfo {author} {\bibfnamefont {P.}~\bibnamefont
  {Kirton}}\ and\ \bibinfo {author} {\bibfnamefont {J.}~\bibnamefont
  {Keeling}},\ }\href {\doibase 10.1088/1367-2630/aaa11d} {\bibfield  {journal}
  {\bibinfo  {journal} {New Journal of Physics}\ }\textbf {\bibinfo {volume}
  {20}},\ \bibinfo {pages} {015009} (\bibinfo {year} {2018})}\BibitemShut
  {NoStop}%
\bibitem [{\citenamefont {Das}\ and\ \citenamefont
  {Sharma}(2022)}]{das2022revisiting}%
  \BibitemOpen
  \bibfield  {author} {\bibinfo {author} {\bibfnamefont {P.}~\bibnamefont
  {Das}}\ and\ \bibinfo {author} {\bibfnamefont {A.}~\bibnamefont {Sharma}},\
  }\href {\doibase 10.1103/PhysRevA.105.033716} {\bibfield  {journal} {\bibinfo
   {journal} {Phys. Rev. A}\ }\textbf {\bibinfo {volume} {105}},\ \bibinfo
  {pages} {033716} (\bibinfo {year} {2022})}\BibitemShut {NoStop}%
\bibitem [{\citenamefont {Dimer}\ \emph
  {et~al.}(2007{\natexlab{b}})\citenamefont {Dimer}, \citenamefont {Estienne},
  \citenamefont {Parkins},\ and\ \citenamefont {Carmichael}}]{Dimer2007}%
  \BibitemOpen
  \bibfield  {author} {\bibinfo {author} {\bibfnamefont {F.}~\bibnamefont
  {Dimer}}, \bibinfo {author} {\bibfnamefont {B.}~\bibnamefont {Estienne}},
  \bibinfo {author} {\bibfnamefont {A.~S.}\ \bibnamefont {Parkins}}, \ and\
  \bibinfo {author} {\bibfnamefont {H.~J.}\ \bibnamefont {Carmichael}},\ }\href
  {\doibase 10.1103/PhysRevA.75.013804} {\bibfield  {journal} {\bibinfo
  {journal} {Phys. Rev. A}\ }\textbf {\bibinfo {volume} {75}},\ \bibinfo
  {pages} {013804} (\bibinfo {year} {2007}{\natexlab{b}})}\BibitemShut
  {NoStop}%
\bibitem [{\citenamefont {Baumann}\ \emph {et~al.}(2010)\citenamefont
  {Baumann}, \citenamefont {Guerlin}, \citenamefont {Brennecke},\ and\
  \citenamefont {Esslinger}}]{baumann_dicke_2010}%
  \BibitemOpen
  \bibfield  {author} {\bibinfo {author} {\bibfnamefont {K.}~\bibnamefont
  {Baumann}}, \bibinfo {author} {\bibfnamefont {C.}~\bibnamefont {Guerlin}},
  \bibinfo {author} {\bibfnamefont {F.}~\bibnamefont {Brennecke}}, \ and\
  \bibinfo {author} {\bibfnamefont {T.}~\bibnamefont {Esslinger}},\ }\href
  {\doibase 10.1038/nature09009} {\bibfield  {journal} {\bibinfo  {journal}
  {Nature}\ }\textbf {\bibinfo {volume} {464}},\ \bibinfo {pages} {1301}
  (\bibinfo {year} {2010})},\ \bibinfo {note} {publisher: Nature Publishing
  Group}\BibitemShut {NoStop}%
\bibitem [{\citenamefont {Léonard}\ \emph {et~al.}(2017)\citenamefont
  {Léonard}, \citenamefont {Morales}, \citenamefont {Zupancic}, \citenamefont
  {Donner},\ and\ \citenamefont {Esslinger}}]{leonard_monitoring_2017}%
  \BibitemOpen
  \bibfield  {author} {\bibinfo {author} {\bibfnamefont {J.}~\bibnamefont
  {Léonard}}, \bibinfo {author} {\bibfnamefont {A.}~\bibnamefont {Morales}},
  \bibinfo {author} {\bibfnamefont {P.}~\bibnamefont {Zupancic}}, \bibinfo
  {author} {\bibfnamefont {T.}~\bibnamefont {Donner}}, \ and\ \bibinfo {author}
  {\bibfnamefont {T.}~\bibnamefont {Esslinger}},\ }\href {\doibase
  10.1126/science.aan2608} {\bibfield  {journal} {\bibinfo  {journal}
  {Science}\ }\textbf {\bibinfo {volume} {358}},\ \bibinfo {pages} {1415}
  (\bibinfo {year} {2017})}\BibitemShut {NoStop}%
\bibitem [{\citenamefont {Guti\'errez-J\'auregui}\ and\ \citenamefont
  {Carmichael}(2018{\natexlab{a}})}]{Guti2018}%
  \BibitemOpen
  \bibfield  {author} {\bibinfo {author} {\bibfnamefont {R.}~\bibnamefont
  {Guti\'errez-J\'auregui}}\ and\ \bibinfo {author} {\bibfnamefont {H.~J.}\
  \bibnamefont {Carmichael}},\ }\href {\doibase 10.1103/PhysRevA.98.023804}
  {\bibfield  {journal} {\bibinfo  {journal} {Phys. Rev. A}\ }\textbf {\bibinfo
  {volume} {98}},\ \bibinfo {pages} {023804} (\bibinfo {year}
  {2018}{\natexlab{a}})}\BibitemShut {NoStop}%
\bibitem [{\citenamefont {Buijsman}\ \emph {et~al.}(2017)\citenamefont
  {Buijsman}, \citenamefont {Gritsev},\ and\ \citenamefont
  {Sprik}}]{buijsman2017nonergodicity}%
  \BibitemOpen
  \bibfield  {author} {\bibinfo {author} {\bibfnamefont {W.}~\bibnamefont
  {Buijsman}}, \bibinfo {author} {\bibfnamefont {V.}~\bibnamefont {Gritsev}}, \
  and\ \bibinfo {author} {\bibfnamefont {R.}~\bibnamefont {Sprik}},\ }\href
  {\doibase 10.1103/PhysRevLett.118.080601} {\bibfield  {journal} {\bibinfo
  {journal} {Phys. Rev. Lett.}\ }\textbf {\bibinfo {volume} {118}},\ \bibinfo
  {pages} {080601} (\bibinfo {year} {2017})}\BibitemShut {NoStop}%
\bibitem [{\citenamefont {Kloc}\ \emph {et~al.}(2017)\citenamefont {Kloc},
  \citenamefont {Str{\'a}nsk{\`y}},\ and\ \citenamefont
  {Cejnar}}]{kloc2017quantum}%
  \BibitemOpen
  \bibfield  {author} {\bibinfo {author} {\bibfnamefont {M.}~\bibnamefont
  {Kloc}}, \bibinfo {author} {\bibfnamefont {P.}~\bibnamefont
  {Str{\'a}nsk{\`y}}}, \ and\ \bibinfo {author} {\bibfnamefont
  {P.}~\bibnamefont {Cejnar}},\ }\href {\doibase
  http://dx.doi.org/10.1016/j.aop.2017.04.005} {\bibfield  {journal} {\bibinfo
  {journal} {Annals of Physics}\ }\textbf {\bibinfo {volume} {382}},\ \bibinfo
  {pages} {85} (\bibinfo {year} {2017})}\BibitemShut {NoStop}%
\bibitem [{\citenamefont {Guti\'errez-J\'auregui}\ and\ \citenamefont
  {Carmichael}(2018{\natexlab{b}})}]{gutierrez2018dissipative}%
  \BibitemOpen
  \bibfield  {author} {\bibinfo {author} {\bibfnamefont {R.}~\bibnamefont
  {Guti\'errez-J\'auregui}}\ and\ \bibinfo {author} {\bibfnamefont {H.~J.}\
  \bibnamefont {Carmichael}},\ }\href {\doibase 10.1103/PhysRevA.98.023804}
  {\bibfield  {journal} {\bibinfo  {journal} {Phys. Rev. A}\ }\textbf {\bibinfo
  {volume} {98}},\ \bibinfo {pages} {023804} (\bibinfo {year}
  {2018}{\natexlab{b}})}\BibitemShut {NoStop}%
\bibitem [{\citenamefont {Shapiro}\ \emph {et~al.}(2020)\citenamefont
  {Shapiro}, \citenamefont {Pogosov},\ and\ \citenamefont
  {Lozovik}}]{shapiro2020universal}%
  \BibitemOpen
  \bibfield  {author} {\bibinfo {author} {\bibfnamefont {D.~S.}\ \bibnamefont
  {Shapiro}}, \bibinfo {author} {\bibfnamefont {W.~V.}\ \bibnamefont
  {Pogosov}}, \ and\ \bibinfo {author} {\bibfnamefont {Y.~E.}\ \bibnamefont
  {Lozovik}},\ }\href {\doibase 10.1103/PhysRevA.102.023703} {\bibfield
  {journal} {\bibinfo  {journal} {Phys. Rev. A}\ }\textbf {\bibinfo {volume}
  {102}},\ \bibinfo {pages} {023703} (\bibinfo {year} {2020})}\BibitemShut
  {NoStop}%
\bibitem [{\citenamefont {Das}\ \emph {et~al.}(2023{\natexlab{a}})\citenamefont
  {Das}, \citenamefont {Bhakuni},\ and\ \citenamefont {Sharma}}]{das2023phase}%
  \BibitemOpen
  \bibfield  {author} {\bibinfo {author} {\bibfnamefont {P.}~\bibnamefont
  {Das}}, \bibinfo {author} {\bibfnamefont {D.~S.}\ \bibnamefont {Bhakuni}}, \
  and\ \bibinfo {author} {\bibfnamefont {A.}~\bibnamefont {Sharma}},\ }\href
  {\doibase 10.1103/PhysRevA.107.043706} {\bibfield  {journal} {\bibinfo
  {journal} {Phys. Rev. A}\ }\textbf {\bibinfo {volume} {107}},\ \bibinfo
  {pages} {043706} (\bibinfo {year} {2023}{\natexlab{a}})}\BibitemShut
  {NoStop}%
\bibitem [{\citenamefont {Hu}\ and\ \citenamefont {Wan}(2021)}]{hu2021out}%
  \BibitemOpen
  \bibfield  {author} {\bibinfo {author} {\bibfnamefont {J.}~\bibnamefont
  {Hu}}\ and\ \bibinfo {author} {\bibfnamefont {S.}~\bibnamefont {Wan}},\
  }\href {\doibase 10.1088/1572-9494/ac256d} {\bibfield  {journal} {\bibinfo
  {journal} {Communications in Theoretical Physics}\ } (\bibinfo {year}
  {2021}),\ 10.1088/1572-9494/ac256d}\BibitemShut {NoStop}%
\bibitem [{\citenamefont {Polkovnikov}\ \emph {et~al.}(2011)\citenamefont
  {Polkovnikov}, \citenamefont {Sengupta}, \citenamefont {Silva},\ and\
  \citenamefont {Vengalattore}}]{RevModPhys.83.863}%
  \BibitemOpen
  \bibfield  {author} {\bibinfo {author} {\bibfnamefont {A.}~\bibnamefont
  {Polkovnikov}}, \bibinfo {author} {\bibfnamefont {K.}~\bibnamefont
  {Sengupta}}, \bibinfo {author} {\bibfnamefont {A.}~\bibnamefont {Silva}}, \
  and\ \bibinfo {author} {\bibfnamefont {M.}~\bibnamefont {Vengalattore}},\
  }\href {\doibase 10.1103/RevModPhys.83.863} {\bibfield  {journal} {\bibinfo
  {journal} {Rev. Mod. Phys.}\ }\textbf {\bibinfo {volume} {83}},\ \bibinfo
  {pages} {863} (\bibinfo {year} {2011})}\BibitemShut {NoStop}%
\bibitem [{\citenamefont {Mori}\ \emph {et~al.}(2018)\citenamefont {Mori},
  \citenamefont {Ikeda}, \citenamefont {Kaminishi},\ and\ \citenamefont
  {Ueda}}]{Mori_2018}%
  \BibitemOpen
  \bibfield  {author} {\bibinfo {author} {\bibfnamefont {T.}~\bibnamefont
  {Mori}}, \bibinfo {author} {\bibfnamefont {T.~N.}\ \bibnamefont {Ikeda}},
  \bibinfo {author} {\bibfnamefont {E.}~\bibnamefont {Kaminishi}}, \ and\
  \bibinfo {author} {\bibfnamefont {M.}~\bibnamefont {Ueda}},\ }\href {\doibase
  10.1088/1361-6455/aabcdf} {\bibfield  {journal} {\bibinfo  {journal} {Journal
  of Physics B: Atomic, Molecular and Optical Physics}\ }\textbf {\bibinfo
  {volume} {51}},\ \bibinfo {pages} {112001} (\bibinfo {year}
  {2018})}\BibitemShut {NoStop}%
\bibitem [{\citenamefont {Serbyn}\ \emph {et~al.}(2021)\citenamefont {Serbyn},
  \citenamefont {Abanin},\ and\ \citenamefont {Papić}}]{serbyn_quantum_2021}%
  \BibitemOpen
  \bibfield  {author} {\bibinfo {author} {\bibfnamefont {M.}~\bibnamefont
  {Serbyn}}, \bibinfo {author} {\bibfnamefont {D.~A.}\ \bibnamefont {Abanin}},
  \ and\ \bibinfo {author} {\bibfnamefont {Z.}~\bibnamefont {Papić}},\ }\href
  {\doibase 10.1038/s41567-021-01230-2} {\bibfield  {journal} {\bibinfo
  {journal} {Nature Physics}\ }\textbf {\bibinfo {volume} {17}},\ \bibinfo
  {pages} {675} (\bibinfo {year} {2021})},\ \bibinfo {note} {publisher: Nature
  Publishing Group}\BibitemShut {NoStop}%
\bibitem [{\citenamefont {D'Alessio}\ \emph {et~al.}(2016)\citenamefont
  {D'Alessio}, \citenamefont {~}, \citenamefont {~}, ,\ and\ \citenamefont
  {Rigol}}]{dalessio_quantum_2016}%
  \BibitemOpen
  \bibfield  {author} {\bibinfo {author} {\bibfnamefont {L.}~\bibnamefont
  {D'Alessio}}, \bibinfo {author} {\bibfnamefont {K.}~\bibnamefont {~},
  \bibfnamefont {Yariv}}, \bibinfo {author} {\bibfnamefont {P.}~\bibnamefont
  {~}, \bibfnamefont {Anatoli}}, , \ and\ \bibinfo {author} {\bibfnamefont
  {M.}~\bibnamefont {Rigol}},\ }\href {\doibase 10.1080/00018732.2016.1198134}
  {\bibfield  {journal} {\bibinfo  {journal} {Advances in Physics}\ }\textbf
  {\bibinfo {volume} {65}},\ \bibinfo {pages} {239} (\bibinfo {year} {2016})},\
  \bibinfo {note} {publisher: Taylor \& Francis \_eprint:
  https://doi.org/10.1080/00018732.2016.1198134}\BibitemShut {NoStop}%
\bibitem [{\citenamefont {Borgonovi}\ \emph {et~al.}(2016)\citenamefont
  {Borgonovi}, \citenamefont {Izrailev}, \citenamefont {Santos},\ and\
  \citenamefont {Zelevinsky}}]{BORGONOVI20161}%
  \BibitemOpen
  \bibfield  {author} {\bibinfo {author} {\bibfnamefont {F.}~\bibnamefont
  {Borgonovi}}, \bibinfo {author} {\bibfnamefont {F.}~\bibnamefont {Izrailev}},
  \bibinfo {author} {\bibfnamefont {L.}~\bibnamefont {Santos}}, \ and\ \bibinfo
  {author} {\bibfnamefont {V.}~\bibnamefont {Zelevinsky}},\ }\href {\doibase
  https://doi.org/10.1016/j.physrep.2016.02.005} {\bibfield  {journal}
  {\bibinfo  {journal} {Physics Reports}\ }\textbf {\bibinfo {volume} {626}},\
  \bibinfo {pages} {1} (\bibinfo {year} {2016})},\ \bibinfo {note} {quantum
  chaos and thermalization in isolated systems of interacting
  particles}\BibitemShut {NoStop}%
\bibitem [{\citenamefont {Bhaseen}\ \emph {et~al.}(2012)\citenamefont
  {Bhaseen}, \citenamefont {Mayoh}, \citenamefont {Simons},\ and\ \citenamefont
  {Keeling}}]{PhysRevA.85.013817}%
  \BibitemOpen
  \bibfield  {author} {\bibinfo {author} {\bibfnamefont {M.~J.}\ \bibnamefont
  {Bhaseen}}, \bibinfo {author} {\bibfnamefont {J.}~\bibnamefont {Mayoh}},
  \bibinfo {author} {\bibfnamefont {B.~D.}\ \bibnamefont {Simons}}, \ and\
  \bibinfo {author} {\bibfnamefont {J.}~\bibnamefont {Keeling}},\ }\href
  {\doibase 10.1103/PhysRevA.85.013817} {\bibfield  {journal} {\bibinfo
  {journal} {Phys. Rev. A}\ }\textbf {\bibinfo {volume} {85}},\ \bibinfo
  {pages} {013817} (\bibinfo {year} {2012})}\BibitemShut {NoStop}%
\bibitem [{\citenamefont {Kirton}\ and\ \citenamefont
  {Keeling}(2017)}]{PhysRevLett.118.123602}%
  \BibitemOpen
  \bibfield  {author} {\bibinfo {author} {\bibfnamefont {P.}~\bibnamefont
  {Kirton}}\ and\ \bibinfo {author} {\bibfnamefont {J.}~\bibnamefont
  {Keeling}},\ }\href {\doibase 10.1103/PhysRevLett.118.123602} {\bibfield
  {journal} {\bibinfo  {journal} {Phys. Rev. Lett.}\ }\textbf {\bibinfo
  {volume} {118}},\ \bibinfo {pages} {123602} (\bibinfo {year}
  {2017})}\BibitemShut {NoStop}%
\bibitem [{\citenamefont {Breuer}\ \emph {et~al.}(2002)\citenamefont {Breuer},
  \citenamefont {Petruccione} \emph {et~al.}}]{breuer2002theory}%
  \BibitemOpen
  \bibfield  {author} {\bibinfo {author} {\bibfnamefont {H.-P.}\ \bibnamefont
  {Breuer}}, \bibinfo {author} {\bibfnamefont {F.}~\bibnamefont {Petruccione}},
   \emph {et~al.},\ }\href@noop {} {\emph {\bibinfo {title} {The theory of open
  quantum systems}}}\ (\bibinfo  {publisher} {Oxford University Press on
  Demand},\ \bibinfo {year} {2002})\BibitemShut {NoStop}%
\bibitem [{\citenamefont {Haga}\ \emph {et~al.}(2021)\citenamefont {Haga},
  \citenamefont {Nakagawa}, \citenamefont {Hamazaki},\ and\ \citenamefont
  {Ueda}}]{haga2021liouvillian}%
  \BibitemOpen
  \bibfield  {author} {\bibinfo {author} {\bibfnamefont {T.}~\bibnamefont
  {Haga}}, \bibinfo {author} {\bibfnamefont {M.}~\bibnamefont {Nakagawa}},
  \bibinfo {author} {\bibfnamefont {R.}~\bibnamefont {Hamazaki}}, \ and\
  \bibinfo {author} {\bibfnamefont {M.}~\bibnamefont {Ueda}},\ }\href {\doibase
  10.1103/PhysRevLett.127.070402} {\bibfield  {journal} {\bibinfo  {journal}
  {Phys. Rev. Lett.}\ }\textbf {\bibinfo {volume} {127}},\ \bibinfo {pages}
  {070402} (\bibinfo {year} {2021})}\BibitemShut {NoStop}%
\bibitem [{\citenamefont {Zhou}\ \emph {et~al.}(2022)\citenamefont {Zhou},
  \citenamefont {Wang},\ and\ \citenamefont {Chen}}]{zhou2022exponential}%
  \BibitemOpen
  \bibfield  {author} {\bibinfo {author} {\bibfnamefont {B.}~\bibnamefont
  {Zhou}}, \bibinfo {author} {\bibfnamefont {X.}~\bibnamefont {Wang}}, \ and\
  \bibinfo {author} {\bibfnamefont {S.}~\bibnamefont {Chen}},\ }\href {\doibase
  10.1103/PhysRevB.106.064203} {\bibfield  {journal} {\bibinfo  {journal}
  {Phys. Rev. B}\ }\textbf {\bibinfo {volume} {106}},\ \bibinfo {pages}
  {064203} (\bibinfo {year} {2022})}\BibitemShut {NoStop}%
\bibitem [{\citenamefont {Prosen}\ and\ \citenamefont
  {Pi{\v{z}}orn}(2008)}]{prosen2008quantum}%
  \BibitemOpen
  \bibfield  {author} {\bibinfo {author} {\bibfnamefont {T.}~\bibnamefont
  {Prosen}}\ and\ \bibinfo {author} {\bibfnamefont {I.}~\bibnamefont
  {Pi{\v{z}}orn}},\ }\href {\doibase 10.1103/PhysRevLett.101.105701} {\bibfield
   {journal} {\bibinfo  {journal} {Phys. Rev. Lett.}\ }\textbf {\bibinfo
  {volume} {101}},\ \bibinfo {pages} {105701} (\bibinfo {year}
  {2008})}\BibitemShut {NoStop}%
\bibitem [{\citenamefont {Kessler}\ \emph {et~al.}(2012)\citenamefont
  {Kessler}, \citenamefont {Giedke}, \citenamefont {Imamoglu}, \citenamefont
  {Yelin}, \citenamefont {Lukin},\ and\ \citenamefont
  {Cirac}}]{kessler2012dissipative}%
  \BibitemOpen
  \bibfield  {author} {\bibinfo {author} {\bibfnamefont {E.~M.}\ \bibnamefont
  {Kessler}}, \bibinfo {author} {\bibfnamefont {G.}~\bibnamefont {Giedke}},
  \bibinfo {author} {\bibfnamefont {A.}~\bibnamefont {Imamoglu}}, \bibinfo
  {author} {\bibfnamefont {S.~F.}\ \bibnamefont {Yelin}}, \bibinfo {author}
  {\bibfnamefont {M.~D.}\ \bibnamefont {Lukin}}, \ and\ \bibinfo {author}
  {\bibfnamefont {J.~I.}\ \bibnamefont {Cirac}},\ }\href {\doibase
  10.1103/PhysRevA.86.012116} {\bibfield  {journal} {\bibinfo  {journal} {Phys.
  Rev. A}\ }\textbf {\bibinfo {volume} {86}},\ \bibinfo {pages} {012116}
  (\bibinfo {year} {2012})}\BibitemShut {NoStop}%
\bibitem [{\citenamefont {Cai}\ and\ \citenamefont
  {Barthel}(2013)}]{cai2013algebraic}%
  \BibitemOpen
  \bibfield  {author} {\bibinfo {author} {\bibfnamefont {Z.}~\bibnamefont
  {Cai}}\ and\ \bibinfo {author} {\bibfnamefont {T.}~\bibnamefont {Barthel}},\
  }\href {\doibase 10.1103/PhysRevLett.111.150403} {\bibfield  {journal}
  {\bibinfo  {journal} {Phys. Rev. Lett.}\ }\textbf {\bibinfo {volume} {111}},\
  \bibinfo {pages} {150403} (\bibinfo {year} {2013})}\BibitemShut {NoStop}%
\bibitem [{\citenamefont {Kastoryano}\ and\ \citenamefont
  {Eisert}(2013)}]{kastoryano2013rapid}%
  \BibitemOpen
  \bibfield  {author} {\bibinfo {author} {\bibfnamefont {M.~J.}\ \bibnamefont
  {Kastoryano}}\ and\ \bibinfo {author} {\bibfnamefont {J.}~\bibnamefont
  {Eisert}},\ }\href@noop {} {\bibfield  {journal} {\bibinfo  {journal}
  {Journal of Mathematical Physics}\ }\textbf {\bibinfo {volume} {54}}
  (\bibinfo {year} {2013})}\BibitemShut {NoStop}%
\bibitem [{\citenamefont {\ifmmode \check{Z}\else
  \v{Z}\fi{}nidari\ifmmode~\check{c}\else
  \v{c}\fi{}}(2015)}]{vznidarivc2015relaxation}%
  \BibitemOpen
  \bibfield  {author} {\bibinfo {author} {\bibfnamefont {M.}~\bibnamefont
  {\ifmmode \check{Z}\else \v{Z}\fi{}nidari\ifmmode~\check{c}\else
  \v{c}\fi{}}},\ }\href {\doibase 10.1103/PhysRevE.92.042143} {\bibfield
  {journal} {\bibinfo  {journal} {Phys. Rev. E}\ }\textbf {\bibinfo {volume}
  {92}},\ \bibinfo {pages} {042143} (\bibinfo {year} {2015})}\BibitemShut
  {NoStop}%
\bibitem [{\citenamefont {Casteels}\ \emph {et~al.}(2017)\citenamefont
  {Casteels}, \citenamefont {Fazio},\ and\ \citenamefont
  {Ciuti}}]{casteels2017critical}%
  \BibitemOpen
  \bibfield  {author} {\bibinfo {author} {\bibfnamefont {W.}~\bibnamefont
  {Casteels}}, \bibinfo {author} {\bibfnamefont {R.}~\bibnamefont {Fazio}}, \
  and\ \bibinfo {author} {\bibfnamefont {C.}~\bibnamefont {Ciuti}},\ }\href
  {\doibase 10.1103/PhysRevA.95.012128} {\bibfield  {journal} {\bibinfo
  {journal} {Phys. Rev. A}\ }\textbf {\bibinfo {volume} {95}},\ \bibinfo
  {pages} {012128} (\bibinfo {year} {2017})}\BibitemShut {NoStop}%
\bibitem [{\citenamefont {Minganti}\ \emph {et~al.}(2018)\citenamefont
  {Minganti}, \citenamefont {Biella}, \citenamefont {Bartolo},\ and\
  \citenamefont {Ciuti}}]{minganti2018spectral}%
  \BibitemOpen
  \bibfield  {author} {\bibinfo {author} {\bibfnamefont {F.}~\bibnamefont
  {Minganti}}, \bibinfo {author} {\bibfnamefont {A.}~\bibnamefont {Biella}},
  \bibinfo {author} {\bibfnamefont {N.}~\bibnamefont {Bartolo}}, \ and\
  \bibinfo {author} {\bibfnamefont {C.}~\bibnamefont {Ciuti}},\ }\href
  {\doibase 10.1103/PhysRevA.98.042118} {\bibfield  {journal} {\bibinfo
  {journal} {Phys. Rev. A}\ }\textbf {\bibinfo {volume} {98}},\ \bibinfo
  {pages} {042118} (\bibinfo {year} {2018})}\BibitemShut {NoStop}%
\bibitem [{\citenamefont {Shibata}\ and\ \citenamefont
  {Katsura}(2019{\natexlab{a}})}]{shibata2019dissipative}%
  \BibitemOpen
  \bibfield  {author} {\bibinfo {author} {\bibfnamefont {N.}~\bibnamefont
  {Shibata}}\ and\ \bibinfo {author} {\bibfnamefont {H.}~\bibnamefont
  {Katsura}},\ }\href {\doibase 10.1103/PhysRevB.99.174303} {\bibfield
  {journal} {\bibinfo  {journal} {Phys. Rev. B}\ }\textbf {\bibinfo {volume}
  {99}},\ \bibinfo {pages} {174303} (\bibinfo {year}
  {2019}{\natexlab{a}})}\BibitemShut {NoStop}%
\bibitem [{\citenamefont {Shibata}\ and\ \citenamefont
  {Katsura}(2019{\natexlab{b}})}]{shibata2019dissipative2}%
  \BibitemOpen
  \bibfield  {author} {\bibinfo {author} {\bibfnamefont {N.}~\bibnamefont
  {Shibata}}\ and\ \bibinfo {author} {\bibfnamefont {H.}~\bibnamefont
  {Katsura}},\ }\href {\doibase 10.1103/PhysRevB.99.224432} {\bibfield
  {journal} {\bibinfo  {journal} {Phys. Rev. B}\ }\textbf {\bibinfo {volume}
  {99}},\ \bibinfo {pages} {224432} (\bibinfo {year}
  {2019}{\natexlab{b}})}\BibitemShut {NoStop}%
\bibitem [{\citenamefont {Shibata}\ and\ \citenamefont
  {Katsura}(2020)}]{shibata2020quantum}%
  \BibitemOpen
  \bibfield  {author} {\bibinfo {author} {\bibfnamefont {N.}~\bibnamefont
  {Shibata}}\ and\ \bibinfo {author} {\bibfnamefont {H.}~\bibnamefont
  {Katsura}},\ }\href {\doibase https://doi.org/10.1093/ptep/ptaa131}
  {\bibfield  {journal} {\bibinfo  {journal} {Progress of Theoretical and
  Experimental Physics}\ }\textbf {\bibinfo {volume} {2020}},\ \bibinfo {pages}
  {12A108} (\bibinfo {year} {2020})}\BibitemShut {NoStop}%
\bibitem [{\citenamefont {Mori}\ and\ \citenamefont
  {Shirai}(2020)}]{mori2020resolving}%
  \BibitemOpen
  \bibfield  {author} {\bibinfo {author} {\bibfnamefont {T.}~\bibnamefont
  {Mori}}\ and\ \bibinfo {author} {\bibfnamefont {T.}~\bibnamefont {Shirai}},\
  }\href {\doibase 10.1103/PhysRevLett.125.230604} {\bibfield  {journal}
  {\bibinfo  {journal} {Phys. Rev. Lett.}\ }\textbf {\bibinfo {volume} {125}},\
  \bibinfo {pages} {230604} (\bibinfo {year} {2020})}\BibitemShut {NoStop}%
\bibitem [{\citenamefont {Mori}\ and\ \citenamefont
  {Shirai}(2023)}]{mori2023symmetrized}%
  \BibitemOpen
  \bibfield  {author} {\bibinfo {author} {\bibfnamefont {T.}~\bibnamefont
  {Mori}}\ and\ \bibinfo {author} {\bibfnamefont {T.}~\bibnamefont {Shirai}},\
  }\href {\doibase 10.1103/PhysRevLett.130.230404} {\bibfield  {journal}
  {\bibinfo  {journal} {Phys. Rev. Lett.}\ }\textbf {\bibinfo {volume} {130}},\
  \bibinfo {pages} {230404} (\bibinfo {year} {2023})}\BibitemShut {NoStop}%
\bibitem [{\citenamefont {Shirai}\ and\ \citenamefont
  {Mori}(2023)}]{shirai2023accelerated}%
  \BibitemOpen
  \bibfield  {author} {\bibinfo {author} {\bibfnamefont {T.}~\bibnamefont
  {Shirai}}\ and\ \bibinfo {author} {\bibfnamefont {T.}~\bibnamefont {Mori}},\
  }\href@noop {} {\bibfield  {journal} {\bibinfo  {journal} {arXiv preprint
  arXiv:2309.03485}\ } (\bibinfo {year} {2023})}\BibitemShut {NoStop}%
\bibitem [{\citenamefont {Mori}(2024)}]{mori2023liouvillian}%
  \BibitemOpen
  \bibfield  {author} {\bibinfo {author} {\bibfnamefont {T.}~\bibnamefont
  {Mori}},\ }\href {\doibase 10.1103/PhysRevB.109.064311} {\bibfield  {journal}
  {\bibinfo  {journal} {Phys. Rev. B}\ }\textbf {\bibinfo {volume} {109}},\
  \bibinfo {pages} {064311} (\bibinfo {year} {2024})}\BibitemShut {NoStop}%
\bibitem [{\citenamefont {Landi}\ \emph {et~al.}(2022)\citenamefont {Landi},
  \citenamefont {Poletti},\ and\ \citenamefont
  {Schaller}}]{landi2022nonequilibrium}%
  \BibitemOpen
  \bibfield  {author} {\bibinfo {author} {\bibfnamefont {G.~T.}\ \bibnamefont
  {Landi}}, \bibinfo {author} {\bibfnamefont {D.}~\bibnamefont {Poletti}}, \
  and\ \bibinfo {author} {\bibfnamefont {G.}~\bibnamefont {Schaller}},\ }\href
  {\doibase 10.1103/RevModPhys.94.045006} {\bibfield  {journal} {\bibinfo
  {journal} {Rev. Mod. Phys.}\ }\textbf {\bibinfo {volume} {94}},\ \bibinfo
  {pages} {045006} (\bibinfo {year} {2022})}\BibitemShut {NoStop}%
\bibitem [{\citenamefont {Yuan}\ \emph {et~al.}(2021)\citenamefont {Yuan},
  \citenamefont {Wang}, \citenamefont {Wang},\ and\ \citenamefont
  {Deng}}]{yuan2021solving}%
  \BibitemOpen
  \bibfield  {author} {\bibinfo {author} {\bibfnamefont {D.}~\bibnamefont
  {Yuan}}, \bibinfo {author} {\bibfnamefont {H.-R.}\ \bibnamefont {Wang}},
  \bibinfo {author} {\bibfnamefont {Z.}~\bibnamefont {Wang}}, \ and\ \bibinfo
  {author} {\bibfnamefont {D.-L.}\ \bibnamefont {Deng}},\ }\href {\doibase
  10.1103/PhysRevLett.126.160401} {\bibfield  {journal} {\bibinfo  {journal}
  {Phys. Rev. Lett.}\ }\textbf {\bibinfo {volume} {126}},\ \bibinfo {pages}
  {160401} (\bibinfo {year} {2021})}\BibitemShut {NoStop}%
\bibitem [{\citenamefont {Ginibre}(1965)}]{ginibre1965statistical}%
  \BibitemOpen
  \bibfield  {author} {\bibinfo {author} {\bibfnamefont {J.}~\bibnamefont
  {Ginibre}},\ }\href@noop {} {\bibfield  {journal} {\bibinfo  {journal}
  {Journal of Mathematical Physics}\ }\textbf {\bibinfo {volume} {6}},\
  \bibinfo {pages} {440} (\bibinfo {year} {1965})}\BibitemShut {NoStop}%
\bibitem [{\citenamefont {Grobe}\ \emph {et~al.}(1988)\citenamefont {Grobe},
  \citenamefont {Haake},\ and\ \citenamefont {Sommers}}]{grobe1988quantum}%
  \BibitemOpen
  \bibfield  {author} {\bibinfo {author} {\bibfnamefont {R.}~\bibnamefont
  {Grobe}}, \bibinfo {author} {\bibfnamefont {F.}~\bibnamefont {Haake}}, \ and\
  \bibinfo {author} {\bibfnamefont {H.-J.}\ \bibnamefont {Sommers}},\ }\href
  {\doibase 10.1103/PhysRevLett.61.1899} {\bibfield  {journal} {\bibinfo
  {journal} {Phys. Rev. Lett.}\ }\textbf {\bibinfo {volume} {61}},\ \bibinfo
  {pages} {1899} (\bibinfo {year} {1988})}\BibitemShut {NoStop}%
\bibitem [{\citenamefont {Markum}\ \emph {et~al.}(1999)\citenamefont {Markum},
  \citenamefont {Pullirsch},\ and\ \citenamefont {Wettig}}]{markum1999non}%
  \BibitemOpen
  \bibfield  {author} {\bibinfo {author} {\bibfnamefont {H.}~\bibnamefont
  {Markum}}, \bibinfo {author} {\bibfnamefont {R.}~\bibnamefont {Pullirsch}}, \
  and\ \bibinfo {author} {\bibfnamefont {T.}~\bibnamefont {Wettig}},\
  }\href@noop {} {\bibfield  {journal} {\bibinfo  {journal} {Physical review
  letters}\ }\textbf {\bibinfo {volume} {83}},\ \bibinfo {pages} {484}
  (\bibinfo {year} {1999})}\BibitemShut {NoStop}%
\bibitem [{\citenamefont {Haake}(1991)}]{haake1991quantum}%
  \BibitemOpen
  \bibfield  {author} {\bibinfo {author} {\bibfnamefont {F.}~\bibnamefont
  {Haake}},\ }\href@noop {} {\emph {\bibinfo {title} {Quantum signatures of
  chaos}}}\ (\bibinfo  {publisher} {Springer},\ \bibinfo {year}
  {1991})\BibitemShut {NoStop}%
\bibitem [{\citenamefont {Hamazaki}\ \emph {et~al.}(2020)\citenamefont
  {Hamazaki}, \citenamefont {Kawabata}, \citenamefont {Kura},\ and\
  \citenamefont {Ueda}}]{hamazaki2020universality}%
  \BibitemOpen
  \bibfield  {author} {\bibinfo {author} {\bibfnamefont {R.}~\bibnamefont
  {Hamazaki}}, \bibinfo {author} {\bibfnamefont {K.}~\bibnamefont {Kawabata}},
  \bibinfo {author} {\bibfnamefont {N.}~\bibnamefont {Kura}}, \ and\ \bibinfo
  {author} {\bibfnamefont {M.}~\bibnamefont {Ueda}},\ }\href {\doibase
  10.1103/PhysRevResearch.2.023286} {\bibfield  {journal} {\bibinfo  {journal}
  {Phys. Rev. Res.}\ }\textbf {\bibinfo {volume} {2}},\ \bibinfo {pages}
  {023286} (\bibinfo {year} {2020})}\BibitemShut {NoStop}%
\bibitem [{\citenamefont {Eckardt}(2017)}]{RevModPhys.89.011004}%
  \BibitemOpen
  \bibfield  {author} {\bibinfo {author} {\bibfnamefont {A.}~\bibnamefont
  {Eckardt}},\ }\href {\doibase 10.1103/RevModPhys.89.011004} {\bibfield
  {journal} {\bibinfo  {journal} {Rev. Mod. Phys.}\ }\textbf {\bibinfo {volume}
  {89}},\ \bibinfo {pages} {011004} (\bibinfo {year} {2017})}\BibitemShut
  {NoStop}%
\bibitem [{\citenamefont {Abanin}\ \emph {et~al.}(2019)\citenamefont {Abanin},
  \citenamefont {Altman}, \citenamefont {Bloch},\ and\ \citenamefont
  {Serbyn}}]{RevModPhys.91.021001}%
  \BibitemOpen
  \bibfield  {author} {\bibinfo {author} {\bibfnamefont {D.~A.}\ \bibnamefont
  {Abanin}}, \bibinfo {author} {\bibfnamefont {E.}~\bibnamefont {Altman}},
  \bibinfo {author} {\bibfnamefont {I.}~\bibnamefont {Bloch}}, \ and\ \bibinfo
  {author} {\bibfnamefont {M.}~\bibnamefont {Serbyn}},\ }\href {\doibase
  10.1103/RevModPhys.91.021001} {\bibfield  {journal} {\bibinfo  {journal}
  {Rev. Mod. Phys.}\ }\textbf {\bibinfo {volume} {91}},\ \bibinfo {pages}
  {021001} (\bibinfo {year} {2019})}\BibitemShut {NoStop}%
\bibitem [{\citenamefont {Nandkishore}\ and\ \citenamefont
  {Huse}(2015)}]{nandkishore_many-body_2015}%
  \BibitemOpen
  \bibfield  {author} {\bibinfo {author} {\bibfnamefont {R.}~\bibnamefont
  {Nandkishore}}\ and\ \bibinfo {author} {\bibfnamefont {D.~A.}\ \bibnamefont
  {Huse}},\ }\href {\doibase 10.1146/annurev-conmatphys-031214-014726}
  {\bibfield  {journal} {\bibinfo  {journal} {Annual Review of Condensed Matter
  Physics}\ }\textbf {\bibinfo {volume} {6}},\ \bibinfo {pages} {15} (\bibinfo
  {year} {2015})},\ \bibinfo {note} {publisher: Annual Reviews}\BibitemShut
  {NoStop}%
\bibitem [{\citenamefont {Abanin}\ \emph {et~al.}(2015)\citenamefont {Abanin},
  \citenamefont {De~Roeck},\ and\ \citenamefont
  {Huveneers}}]{PhysRevLett.115.256803}%
  \BibitemOpen
  \bibfield  {author} {\bibinfo {author} {\bibfnamefont {D.~A.}\ \bibnamefont
  {Abanin}}, \bibinfo {author} {\bibfnamefont {W.}~\bibnamefont {De~Roeck}}, \
  and\ \bibinfo {author} {\bibfnamefont {F.~m.~c.}\ \bibnamefont {Huveneers}},\
  }\href {\doibase 10.1103/PhysRevLett.115.256803} {\bibfield  {journal}
  {\bibinfo  {journal} {Phys. Rev. Lett.}\ }\textbf {\bibinfo {volume} {115}},\
  \bibinfo {pages} {256803} (\bibinfo {year} {2015})}\BibitemShut {NoStop}%
\bibitem [{\citenamefont {Beatrez}\ \emph {et~al.}(2021)\citenamefont
  {Beatrez}, \citenamefont {Janes}, \citenamefont {Akkiraju}, \citenamefont
  {Pillai}, \citenamefont {Oddo}, \citenamefont {Reshetikhin}, \citenamefont
  {Druga}, \citenamefont {McAllister}, \citenamefont {Elo}, \citenamefont
  {Gilbert}, \citenamefont {Suter},\ and\ \citenamefont
  {Ajoy}}]{PhysRevLett.127.170603}%
  \BibitemOpen
  \bibfield  {author} {\bibinfo {author} {\bibfnamefont {W.}~\bibnamefont
  {Beatrez}}, \bibinfo {author} {\bibfnamefont {O.}~\bibnamefont {Janes}},
  \bibinfo {author} {\bibfnamefont {A.}~\bibnamefont {Akkiraju}}, \bibinfo
  {author} {\bibfnamefont {A.}~\bibnamefont {Pillai}}, \bibinfo {author}
  {\bibfnamefont {A.}~\bibnamefont {Oddo}}, \bibinfo {author} {\bibfnamefont
  {P.}~\bibnamefont {Reshetikhin}}, \bibinfo {author} {\bibfnamefont
  {E.}~\bibnamefont {Druga}}, \bibinfo {author} {\bibfnamefont
  {M.}~\bibnamefont {McAllister}}, \bibinfo {author} {\bibfnamefont
  {M.}~\bibnamefont {Elo}}, \bibinfo {author} {\bibfnamefont {B.}~\bibnamefont
  {Gilbert}}, \bibinfo {author} {\bibfnamefont {D.}~\bibnamefont {Suter}}, \
  and\ \bibinfo {author} {\bibfnamefont {A.}~\bibnamefont {Ajoy}},\ }\href
  {\doibase 10.1103/PhysRevLett.127.170603} {\bibfield  {journal} {\bibinfo
  {journal} {Phys. Rev. Lett.}\ }\textbf {\bibinfo {volume} {127}},\ \bibinfo
  {pages} {170603} (\bibinfo {year} {2021})}\BibitemShut {NoStop}%
\bibitem [{\citenamefont {Saha}\ and\ \citenamefont
  {Bhattacharyya}(2023)}]{PhysRevA.107.022206}%
  \BibitemOpen
  \bibfield  {author} {\bibinfo {author} {\bibfnamefont {S.}~\bibnamefont
  {Saha}}\ and\ \bibinfo {author} {\bibfnamefont {R.}~\bibnamefont
  {Bhattacharyya}},\ }\href {\doibase 10.1103/PhysRevA.107.022206} {\bibfield
  {journal} {\bibinfo  {journal} {Phys. Rev. A}\ }\textbf {\bibinfo {volume}
  {107}},\ \bibinfo {pages} {022206} (\bibinfo {year} {2023})}\BibitemShut
  {NoStop}%
\bibitem [{\citenamefont {Das}\ \emph {et~al.}(2023{\natexlab{b}})\citenamefont
  {Das}, \citenamefont {Bhakuni}, \citenamefont {Santos},\ and\ \citenamefont
  {Sharma}}]{das2023periodically}%
  \BibitemOpen
  \bibfield  {author} {\bibinfo {author} {\bibfnamefont {P.}~\bibnamefont
  {Das}}, \bibinfo {author} {\bibfnamefont {D.~S.}\ \bibnamefont {Bhakuni}},
  \bibinfo {author} {\bibfnamefont {L.~F.}\ \bibnamefont {Santos}}, \ and\
  \bibinfo {author} {\bibfnamefont {A.}~\bibnamefont {Sharma}},\ }\href
  {\doibase 10.1103/PhysRevA.108.063716} {\bibfield  {journal} {\bibinfo
  {journal} {Phys. Rev. A}\ }\textbf {\bibinfo {volume} {108}},\ \bibinfo
  {pages} {063716} (\bibinfo {year} {2023}{\natexlab{b}})}\BibitemShut
  {NoStop}%
\bibitem [{\citenamefont {Fleckenstein}\ and\ \citenamefont
  {Bukov}(2021)}]{PhysRevB.103.L140302}%
  \BibitemOpen
  \bibfield  {author} {\bibinfo {author} {\bibfnamefont {C.}~\bibnamefont
  {Fleckenstein}}\ and\ \bibinfo {author} {\bibfnamefont {M.}~\bibnamefont
  {Bukov}},\ }\href {\doibase 10.1103/PhysRevB.103.L140302} {\bibfield
  {journal} {\bibinfo  {journal} {Phys. Rev. B}\ }\textbf {\bibinfo {volume}
  {103}},\ \bibinfo {pages} {L140302} (\bibinfo {year} {2021})}\BibitemShut
  {NoStop}%
\bibitem [{\citenamefont {Nandy}\ \emph {et~al.}(2017)\citenamefont {Nandy},
  \citenamefont {Sen},\ and\ \citenamefont {Sen}}]{nandy2017aperiodically}%
  \BibitemOpen
  \bibfield  {author} {\bibinfo {author} {\bibfnamefont {S.}~\bibnamefont
  {Nandy}}, \bibinfo {author} {\bibfnamefont {A.}~\bibnamefont {Sen}}, \ and\
  \bibinfo {author} {\bibfnamefont {D.}~\bibnamefont {Sen}},\ }\href {\doibase
  10.1103/PhysRevX.7.031034} {\bibfield  {journal} {\bibinfo  {journal} {Phys.
  Rev. X}\ }\textbf {\bibinfo {volume} {7}},\ \bibinfo {pages} {031034}
  (\bibinfo {year} {2017})}\BibitemShut {NoStop}%
\bibitem [{\citenamefont {Mukherjee}\ \emph {et~al.}(2020)\citenamefont
  {Mukherjee}, \citenamefont {Sen}, \citenamefont {Sen},\ and\ \citenamefont
  {Sengupta}}]{mukherjee2020restoring}%
  \BibitemOpen
  \bibfield  {author} {\bibinfo {author} {\bibfnamefont {B.}~\bibnamefont
  {Mukherjee}}, \bibinfo {author} {\bibfnamefont {A.}~\bibnamefont {Sen}},
  \bibinfo {author} {\bibfnamefont {D.}~\bibnamefont {Sen}}, \ and\ \bibinfo
  {author} {\bibfnamefont {K.}~\bibnamefont {Sengupta}},\ }\href {\doibase
  10.1103/PhysRevB.102.014301} {\bibfield  {journal} {\bibinfo  {journal}
  {Phys. Rev. B}\ }\textbf {\bibinfo {volume} {102}},\ \bibinfo {pages}
  {014301} (\bibinfo {year} {2020})}\BibitemShut {NoStop}%
\bibitem [{\citenamefont {Zhao}\ \emph {et~al.}(2021)\citenamefont {Zhao},
  \citenamefont {Mintert}, \citenamefont {Moessner},\ and\ \citenamefont
  {Knolle}}]{zhao2021random}%
  \BibitemOpen
  \bibfield  {author} {\bibinfo {author} {\bibfnamefont {H.}~\bibnamefont
  {Zhao}}, \bibinfo {author} {\bibfnamefont {F.}~\bibnamefont {Mintert}},
  \bibinfo {author} {\bibfnamefont {R.}~\bibnamefont {Moessner}}, \ and\
  \bibinfo {author} {\bibfnamefont {J.}~\bibnamefont {Knolle}},\ }\href
  {\doibase 10.1103/PhysRevLett.126.040601} {\bibfield  {journal} {\bibinfo
  {journal} {Phys. Rev. Lett.}\ }\textbf {\bibinfo {volume} {126}},\ \bibinfo
  {pages} {040601} (\bibinfo {year} {2021})}\BibitemShut {NoStop}%
\bibitem [{\citenamefont {Zhao}\ \emph {et~al.}(2022)\citenamefont {Zhao},
  \citenamefont {Mintert}, \citenamefont {Knolle},\ and\ \citenamefont
  {Moessner}}]{zhao2022localization}%
  \BibitemOpen
  \bibfield  {author} {\bibinfo {author} {\bibfnamefont {H.}~\bibnamefont
  {Zhao}}, \bibinfo {author} {\bibfnamefont {F.}~\bibnamefont {Mintert}},
  \bibinfo {author} {\bibfnamefont {J.}~\bibnamefont {Knolle}}, \ and\ \bibinfo
  {author} {\bibfnamefont {R.}~\bibnamefont {Moessner}},\ }\href {\doibase
  10.1103/PhysRevB.105.L220202} {\bibfield  {journal} {\bibinfo  {journal}
  {Phys. Rev. B}\ }\textbf {\bibinfo {volume} {105}},\ \bibinfo {pages}
  {L220202} (\bibinfo {year} {2022})}\BibitemShut {NoStop}%
\bibitem [{\citenamefont {Tiwari}\ \emph {et~al.}(2024)\citenamefont {Tiwari},
  \citenamefont {Bhakuni},\ and\ \citenamefont {Sharma}}]{tiwari2024dynamical}%
  \BibitemOpen
  \bibfield  {author} {\bibinfo {author} {\bibfnamefont {V.}~\bibnamefont
  {Tiwari}}, \bibinfo {author} {\bibfnamefont {D.~S.}\ \bibnamefont {Bhakuni}},
  \ and\ \bibinfo {author} {\bibfnamefont {A.}~\bibnamefont {Sharma}},\ }\href
  {\doibase 10.1103/PhysRevB.109.L161104} {\bibfield  {journal} {\bibinfo
  {journal} {Phys. Rev. B}\ }\textbf {\bibinfo {volume} {109}},\ \bibinfo
  {pages} {L161104} (\bibinfo {year} {2024})}\BibitemShut {NoStop}%
\bibitem [{\citenamefont {Tiwari}\ \emph {et~al.}(2025)\citenamefont {Tiwari},
  \citenamefont {Bhakuni},\ and\ \citenamefont
  {Sharma}}]{tiwari2024periodically}%
  \BibitemOpen
  \bibfield  {author} {\bibinfo {author} {\bibfnamefont {V.}~\bibnamefont
  {Tiwari}}, \bibinfo {author} {\bibfnamefont {D.~S.}\ \bibnamefont {Bhakuni}},
  \ and\ \bibinfo {author} {\bibfnamefont {A.}~\bibnamefont {Sharma}},\ }\href
  {\doibase 10.1103/PhysRevB.111.205109} {\bibfield  {journal} {\bibinfo
  {journal} {Phys. Rev. B}\ }\textbf {\bibinfo {volume} {111}},\ \bibinfo
  {pages} {205109} (\bibinfo {year} {2025})}\BibitemShut {NoStop}%
\bibitem [{\citenamefont {Prasad}\ \emph {et~al.}(2022)\citenamefont {Prasad},
  \citenamefont {Yadalam}, \citenamefont {Aron},\ and\ \citenamefont
  {Kulkarni}}]{prasad2022dissipative}%
  \BibitemOpen
  \bibfield  {author} {\bibinfo {author} {\bibfnamefont {M.}~\bibnamefont
  {Prasad}}, \bibinfo {author} {\bibfnamefont {H.~K.}\ \bibnamefont {Yadalam}},
  \bibinfo {author} {\bibfnamefont {C.}~\bibnamefont {Aron}}, \ and\ \bibinfo
  {author} {\bibfnamefont {M.}~\bibnamefont {Kulkarni}},\ }\href {\doibase
  10.1103/PhysRevA.105.L050201} {\bibfield  {journal} {\bibinfo  {journal}
  {Phys. Rev. A}\ }\textbf {\bibinfo {volume} {105}},\ \bibinfo {pages}
  {L050201} (\bibinfo {year} {2022})}\BibitemShut {NoStop}%
\bibitem [{\citenamefont {Saha}\ and\ \citenamefont
  {Bhattacharyya}(2024)}]{saha2024prethermalization}%
  \BibitemOpen
  \bibfield  {author} {\bibinfo {author} {\bibfnamefont {S.}~\bibnamefont
  {Saha}}\ and\ \bibinfo {author} {\bibfnamefont {R.}~\bibnamefont
  {Bhattacharyya}},\ }\href {\doibase 10.1088/1742-5468/ad1d56} {\bibfield
  {journal} {\bibinfo  {journal} {Journal of Statistical Mechanics: Theory and
  Experiment}\ }\textbf {\bibinfo {volume} {2024}},\ \bibinfo {pages} {023103}
  (\bibinfo {year} {2024})}\BibitemShut {NoStop}%
\bibitem [{\citenamefont {Villase\~nor}\ \emph {et~al.}(2024)\citenamefont
  {Villase\~nor}, \citenamefont {Santos},\ and\ \citenamefont
  {Barberis-Blostein}}]{villasenor2024breakdown}%
  \BibitemOpen
  \bibfield  {author} {\bibinfo {author} {\bibfnamefont {D.}~\bibnamefont
  {Villase\~nor}}, \bibinfo {author} {\bibfnamefont {L.~F.}\ \bibnamefont
  {Santos}}, \ and\ \bibinfo {author} {\bibfnamefont {P.}~\bibnamefont
  {Barberis-Blostein}},\ }\href {\doibase 10.1103/PhysRevLett.133.240404}
  {\bibfield  {journal} {\bibinfo  {journal} {Phys. Rev. Lett.}\ }\textbf
  {\bibinfo {volume} {133}},\ \bibinfo {pages} {240404} (\bibinfo {year}
  {2024})}\BibitemShut {NoStop}%
\bibitem [{\citenamefont {Thue}(1906)}]{thue1906uber}%
  \BibitemOpen
  \bibfield  {author} {\bibinfo {author} {\bibfnamefont {A.}~\bibnamefont
  {Thue}},\ }\href@noop {} {\bibfield  {journal} {\bibinfo  {journal} {Norske
  Vid Selsk. Skr. I Mat-Nat Kl.(Christiana)}\ }\textbf {\bibinfo {volume}
  {7}},\ \bibinfo {pages} {1} (\bibinfo {year} {1906})}\BibitemShut {NoStop}%
\bibitem [{\citenamefont {He}\ \emph {et~al.}(2025)\citenamefont {He},
  \citenamefont {Ye}, \citenamefont {Gong}, \citenamefont {Yao}, \citenamefont
  {Liu}, \citenamefont {Murch}, \citenamefont {Yao},\ and\ \citenamefont
  {Zu}}]{he2025experimental}%
  \BibitemOpen
  \bibfield  {author} {\bibinfo {author} {\bibfnamefont {G.}~\bibnamefont
  {He}}, \bibinfo {author} {\bibfnamefont {B.}~\bibnamefont {Ye}}, \bibinfo
  {author} {\bibfnamefont {R.}~\bibnamefont {Gong}}, \bibinfo {author}
  {\bibfnamefont {C.}~\bibnamefont {Yao}}, \bibinfo {author} {\bibfnamefont
  {Z.}~\bibnamefont {Liu}}, \bibinfo {author} {\bibfnamefont {K.~W.}\
  \bibnamefont {Murch}}, \bibinfo {author} {\bibfnamefont {N.~Y.}\ \bibnamefont
  {Yao}}, \ and\ \bibinfo {author} {\bibfnamefont {C.}~\bibnamefont {Zu}},\
  }\href {\doibase 10.1103/PhysRevX.15.011055} {\bibfield  {journal} {\bibinfo
  {journal} {Phys. Rev. X}\ }\textbf {\bibinfo {volume} {15}},\ \bibinfo
  {pages} {011055} (\bibinfo {year} {2025})}\BibitemShut {NoStop}%
\bibitem [{\citenamefont {Berry}\ and\ \citenamefont
  {Tabor}(1977)}]{berry1977level}%
  \BibitemOpen
  \bibfield  {author} {\bibinfo {author} {\bibfnamefont {M.~V.}\ \bibnamefont
  {Berry}}\ and\ \bibinfo {author} {\bibfnamefont {M.}~\bibnamefont {Tabor}},\
  }\href {\doibase https://doi.org/10.1098/rspa.1977.0140} {\bibfield
  {journal} {\bibinfo  {journal} {Proceedings of the Royal Society of London.
  A. Mathematical and Physical Sciences}\ }\textbf {\bibinfo {volume} {356}},\
  \bibinfo {pages} {375} (\bibinfo {year} {1977})}\BibitemShut {NoStop}%
\bibitem [{\citenamefont {Bohr}\ and\ \citenamefont
  {Mottelson}(1969)}]{bohr1969vol}%
  \BibitemOpen
  \bibfield  {author} {\bibinfo {author} {\bibfnamefont {A.}~\bibnamefont
  {Bohr}}\ and\ \bibinfo {author} {\bibfnamefont {B.}~\bibnamefont
  {Mottelson}},\ }\href@noop {} {\enquote {\bibinfo {title} {Vol. 1, appendix
  2-d, nuclear structure},}\ } (\bibinfo {year} {1969})\BibitemShut {NoStop}%
\bibitem [{\citenamefont {Bohigas}\ \emph {et~al.}(1984)\citenamefont
  {Bohigas}, \citenamefont {Giannoni},\ and\ \citenamefont
  {Schmit}}]{bohigas1984characterization}%
  \BibitemOpen
  \bibfield  {author} {\bibinfo {author} {\bibfnamefont {O.}~\bibnamefont
  {Bohigas}}, \bibinfo {author} {\bibfnamefont {M.~J.}\ \bibnamefont
  {Giannoni}}, \ and\ \bibinfo {author} {\bibfnamefont {C.}~\bibnamefont
  {Schmit}},\ }\href {\doibase 10.1103/PhysRevLett.52.1} {\bibfield  {journal}
  {\bibinfo  {journal} {Phys. Rev. Lett.}\ }\textbf {\bibinfo {volume} {52}},\
  \bibinfo {pages} {1} (\bibinfo {year} {1984})}\BibitemShut {NoStop}%
\bibitem [{\citenamefont {Akemann}\ \emph {et~al.}(2019)\citenamefont
  {Akemann}, \citenamefont {Kieburg}, \citenamefont {Mielke},\ and\
  \citenamefont {Prosen}}]{akemann2019universal}%
  \BibitemOpen
  \bibfield  {author} {\bibinfo {author} {\bibfnamefont {G.}~\bibnamefont
  {Akemann}}, \bibinfo {author} {\bibfnamefont {M.}~\bibnamefont {Kieburg}},
  \bibinfo {author} {\bibfnamefont {A.}~\bibnamefont {Mielke}}, \ and\ \bibinfo
  {author} {\bibfnamefont {T.}~\bibnamefont {Prosen}},\ }\href {\doibase
  10.1103/PhysRevLett.123.254101} {\bibfield  {journal} {\bibinfo  {journal}
  {Phys. Rev. Lett.}\ }\textbf {\bibinfo {volume} {123}},\ \bibinfo {pages}
  {254101} (\bibinfo {year} {2019})}\BibitemShut {NoStop}%
\bibitem [{\citenamefont {S{\'a}}\ \emph {et~al.}(2020)\citenamefont {S{\'a}},
  \citenamefont {Ribeiro},\ and\ \citenamefont {Prosen}}]{sa2020complex}%
  \BibitemOpen
  \bibfield  {author} {\bibinfo {author} {\bibfnamefont {L.}~\bibnamefont
  {S{\'a}}}, \bibinfo {author} {\bibfnamefont {P.}~\bibnamefont {Ribeiro}}, \
  and\ \bibinfo {author} {\bibfnamefont {T.}~\bibnamefont {Prosen}},\ }\href
  {\doibase 10.1103/PhysRevX.10.021019} {\bibfield  {journal} {\bibinfo
  {journal} {Phys. Rev. X}\ }\textbf {\bibinfo {volume} {10}},\ \bibinfo
  {pages} {021019} (\bibinfo {year} {2020})}\BibitemShut {NoStop}%
\end{thebibliography}%
\end{document}